\documentclass[aps,a4paper,showpacs,preprintnumbers,amsmath,amssymb,
 floatfix]{revtex4}
\usepackage{dcolumn}
\usepackage{bm}
\usepackage{amsmath,amssymb}
\usepackage{booktabs,subfigure}
\usepackage{graphicx,psfrag}
\usepackage{epsfig}
\usepackage{color} 
\usepackage{rotating}
\usepackage{axodraw}
\usepackage{slashed}
\usepackage{geometry}
\usepackage{rotating}
\allowdisplaybreaks[3]
\newcommand{\cw}[1][{}]{\ensuremath{\cos^{#1} \theta_{w}}}
\newcommand{\sw}[1][{}]{\ensuremath{\sin^{#1} \theta_{w}}}
\newcommand{\tw}[1][{}]{\ensuremath{\tan^{#1} \theta_{w}}}

\newcommand{\M}{\ensuremath{\mathcal{M}}}

\newcommand{\Backgroundnu}[1][{}]{\ensuremath{e^+e^-  \to \nu \bar\nu \gamma}#1}

\newcommand{\signal}[1][{}]{\ensuremath{e^+e^-\rightarrow \tilde\chi^0_1 \tilde\chi^0_1\gamma}#1}

\newcommand{\GeV}{\ensuremath{~\mathrm{GeV}}}
\newcommand{\fb}{\ensuremath{~\mathrm{fb}}}

\numberwithin{equation}{section} 
\def\lsim{\raise0.3ex\hbox{$\;<$\kern-0.75em\raise-1.1ex\hbox{$\sim\;$}}}
\def\gsim{\raise0.3ex\hbox{$\;>$\kern-0.75em\raise-1.1ex\hbox{$\sim\;$}}}
\begin{document}
\title{ Radiative neutralino production in low energy supersymmetric
Models. II. \\ The case of beam polarization}

\author{P. N. Pandita$^{1}$
and %
Chandradew Sharma$^2$}
%
\affiliation{
            $^1$ Department of Physics, North Eastern Hill University,
                 Shillong 793 022, India\\
            $^2$  Department of Physics, Birla Institute of Technology and Science, Pilani - K. K. 
                 Birla Goa Campus, Goa 403 726, India}
%
\thispagestyle{myheadings}

 
\begin{abstract}
\noindent
We study the production of the lightest neutralinos in the radiative process
$e^+e^- \to \tilde\chi^0_1 \tilde\chi^0_1\gamma$ in low energy 
supersymmetric models for the International Linear Collider energies
with longitudinally polarized electron and positron beams. 
For this purpose we  consider the case of nonminimal supersymmetric
standard model as well as the  case of minimal supersymmetric standard model.
At the first stage of  a linear collider,  with $\sqrt{s} =500$ GeV,
the radiative production of the lightest neutralinos
may be  a viable channel to study
supersymmetric partners of the Standard Model particles,
especially if the other supersymmetric particles are too
heavy to  to be pair-produced.
We consider in detail
the  effect of beam polarization  on the production cross section. 
We  compare and contrast the dependence of the signal cross
section on the parameters of the neutralino sector of the nonminimal and 
minimal supersymmetric standard model when the electron and 
positron beams are longitudinally polarized.
In order to assess the feasibility of experimentally observing the 
radiative neutralino production process, we consider the background to 
this process coming from the
Standard Model process  $e^+e^- \to \nu \bar\nu \gamma$ with longitudinally 
polarized electron and positron beams. We also consider the supersymmetric
background to the radiative neutralino production process coming from
the radiative production of the scalar partners of the 
neutrinos~(sneutrinos)
$e^+e^- \to \tilde\nu \tilde\nu^\ast \gamma$, with longitudinally
polarized beams. This process can be a a background 
to the radiative neutralino production when the sneutrinos decay invisibly.
\end{abstract}
\pacs{11.30.Pb, 12.60.Jv, 14.80.Ly}
\maketitle

\section{Introduction}
\label{sec:intro}
Supersymmetry is at present one of the most favored  ideas for physics 
beyond the  Standard Model~(SM)~\cite{Wess:1992cp, kaul}.
A particularly attractive implementation of the idea of supersymmetry
is the Minimal Supersymmetric Standard Model~(MSSM) obtained by 
introducing the supersymmetric partners of the  of the SM states, and 
introducing an additional Higgs doublet, with opposite hypercharge to that 
of the SM Higgs doublet, in order to cancel the gauge 
anomalies and generate masses for all the fermions of the Standard 
Model~\cite{Nilles:1983ge, Drees:2004jm}. 
Supersymmetry must obviously be a broken symmetry.
In order for broken supersymmetry to be 
effective in protecting the weak scale against large radiative corrections, 
the supersymmetric partners of the SM particles should have masses  of the 
order of a few hundred GeV.  Their  discovery is one of the main goals 
of present and future 
accelerators.  In particular, a  $e^+e^-$ linear collider 
with a high luminosity ${\mathcal L}=500$ fb$^{-1}$,  and a center-of-mass
energy of $\sqrt s = 500$ GeV in the first stage,  will be
an important tool in determining the parameters of the
low energy supersymmetric model with a high 
precision~\cite{Aguilar-Saavedra:2001rg,Abe:2001nn,Abe:2001gc, 
Weiglein:2004hn,Aguilar-Saavedra:2005pw}.  Furthermore, polarization
of the electron (and positron) beam can enhance the 
capability of such a linear collider~\cite{Moortgat-Pick:2005cw}
in unravelling the structure of the underlying supersymmetric
model.

In the minimal supersymmetric standard model the fermionic partners
of the two Higgs doublets~($H_1, H_2$) mix with the fermionic partners
of the gauge bosons to produce four neutralino states $\tilde \chi^0_i$,
$i = 1, 2, 3, 4$, and two chargino states  $\tilde \chi^{\pm}_j$,
$j = 1, 2.$  
In the MSSM the lightest neutralino is favored to be the lightest 
supersymmetric particle, and assuming, $R$-parity~($R_p$) conservation, 
is absolutely stable.
The neutralino states of the minimal supersymmetric
standard model with $R_p$ conservation have been studied in great detail, 
because
the lightest neutralino, being the lightest supersymmetric 
particle~(LSP), is the end product 
of any process involving supersymmetric particle in the final state.

There are alternatives to the MSSM, an elegant one being the model with an 
additional chiral electroweak gauge singlet Higgs superfield $S$
which couples to the two Higgs doublet
superfields $H_1$ and $H_2$ via a dimensionless trilinear term
$\lambda H_1 H_2 S$ in the superpotential. This model can solve the 
the $\mu$ problem of the MSSM in a natural manner. When the scalar component
of the singlet superfield $S$ obtains a vacuum expectation value, a
bilinear term $\lambda H_1 H_2 <S>$ involving the two Higgs doublets
is naturally generated.  Furthermore, when this scalar component of
the chiral singlet superfield $S$ acquires a vacuum expectation value
of the order of the $SU(2)_L \times U(1)_Y$ breaking scale, it gives
rise to an effective value of 
$\mu$~($\mu_{eff} \equiv \lambda <S> = \lambda x$) of the
order of the electroweak scale.  However, the inclusion of the singlet
superfield leads to additional trilinear superpotential coupling
$(\kappa/ 3 ) S^3$ in the model, the so called nonminimal, or
next-to-minimal~\cite{fayet, ellis89, drees89, nmssm1, nmssm2, nmssm3, nmssm4},
supersymmetric standard model~(NMSSM).  The absence of $H_1 H_2$ term,
and the absence of tadpole and mass couplings, $S$ and $ S^2$ in the
NMSSM is made natural by postulating a suitable discrete 
symmetry~\cite{Chemtob:2006ur, Chemtob:2007rg}.
The NMSSM is attractive on account of the simple resolution it 
offers to the $\mu$ problem, and of the scale invariance of its 
classical action in the supersymmetric limit~\cite{Ellwanger:2009dp}. 
Since no dimensional 
supersymmetric parameters are present in the superpotential of NMSSM, 
it is the simplest supersymmetric extension of the Standard Model in 
which the electroweak scale originates from the supersymmetry breaking 
scale only.  Its enlarged Higgs sector may help in relaxing the 
fine-tuning and little hierarchy problems of the MSSM, 
thereby opening new perspectives for the Higgs boson searches at high 
energy colliders~\cite{ellwang04,  moort05}, and for dark
matter searches~\cite{gunion05}. In the nonminimal supersymmetric standard 
model the mixing of fermionic partners of Higgs and gauge 
bosons~\cite{pnp1, pnp2, choi04}  produces five 
neutralino states  $\tilde \chi^0_i$,
$i = 1, 2, 3, 4, 5$, and two chargino states  $\tilde \chi^{\pm}_j$,
$j = 1, 2.$  Furthermore, because of the presence of the
fermionic partner of the singlet Higgs boson, the neutralino states can 
have an admixture of this $SU(2)_L \times U(1)_Y$ singlet fermion, 
thereby affecting the phenomenology of the
neutralinos in the nonminimal supersymmetric standard model.

The lightest neutralino state~($\tilde\chi_1^0$) of MSSM or NMSSM, being 
typically the LSP,  is 
stable and therefore, a possible dark matter 
candidate~\cite{Goldberg:1983nd, Ellis:1983ew}.
Since the neutralinos are among the lightest particles
in low energy supersymmetric models, 
they are expected to be the first states to
be produced at the colliding beam experiments.  
At an electron-positron collider, such as the  International
Linear Collider~(ILC), the lightest neutralino can be 
produced in pairs
\begin{equation}
e^+ + e^-\to\tilde\chi_1^0 + \tilde\chi_1^0\,.
\label{pairneuts}
\end{equation}
This process  proceeds via $Z$ boson and selectron 
exchange~\cite{Bartl:1986hp, Ellis}.
In collider experiments the  lightest neutralino 
escapes detection. In  such a situation the  production of the lightest
neutralino pair (\ref{pairneuts})
is invisible.  Therefore, we must look for the signature of 
neutralinos in the radiative process
\begin{equation}
e^- + e^+ \to\tilde\chi_1^0 + \tilde\chi_1^0 + \gamma.
\label{radiative}
\end{equation}
Despite this process being   suppressed 
by the square of the electromagnetic coupling, it might be the first 
process where the lightest supersymmetric states could  be observed at 
the $e^+ e^-$ colliders.  The signal of the radiative process (\ref{radiative})
is a single high energy photon with the missing energy  carried away 
by the neutralinos.  The process~(\ref{radiative}) has been
studied in detail in the minimal supersymmetric  
standard model~\cite{Fayet:1982ky,Ellis:1982zz, Grassie:1983kq,
Kobayashi:1984wu,Ware:1984kq,Bento:1985in, Chen:1987ux,Kon:1987gi,
Ambrosanio:1995it, Choi:1999bs, Datta:1996ur}.
Some of these studies underline the importance of 
longitudinal~\cite{ Choi:1999bs}, 
and even transverse beam polarizations. On the other hand, the signature 
``photon plus missing energy,'' that arises in the process 
(\ref{radiative}) has been studied
in detail by different LEP collaborations~\cite{Heister:2002ut,
Abdallah:2003np, Achard:2003tx, Abbiendi:2002vz,Abbiendi:2000hh}.  
Furthermore,  the radiative neutrino process  
$e^+e^- \to \nu \bar\nu \gamma$ in the SM 
is the leading process with this signature, for which
the cross section depends on the number $N_\nu$ of light neutrino
species~\cite{Gaemers:1978fe}.  The LEP collaborations have found no 
deviations from the SM prediction, and, therefore,  only bounds on 
the masses of supersymmetric particles have 
been set~\cite{Heister:2002ut,Abdallah:2003np,Achard:2003tx,Abbiendi:2000hh}.  
For a review of the experimental situation, see Ref.~\cite{Gataullin:2003sy}.

Most of the theoretical studies on radiative neutralino
production in the literature have been carried out in the framework of the
minimal supersymmetric standard model. 
This includes calculations relevant to ILC with a high center-of-mass energy, 
high luminosity and longitudinally polarized beams, as well as study of the SM 
background from the radiative neutrino production 
\begin{equation}
e^+e^- \to  \nu + \bar\nu + \gamma, 
\label{radiativenu} 
\end{equation}
and the supersymmetric background from radiative sneutrino 
production  
\begin{equation}
e^+e^- \to \tilde\nu + \tilde\nu^\ast + \gamma.
\label{radiativesnu}
\end{equation}
The discovery potential of ILC may be significantly 
enhanced~\cite{Dreiner:2006sb} if  both beams are polarized, 
particularly if  other SUSY states 
like heavier neutralino, chargino or
even slepton pairs are too heavy to be produced at the first stage of
the ILC at $\sqrt s = 500$~GeV.  

In a previous paper~(referred to as paper I)
we have carried out a detailed study of  the 
radiative process~(\ref{radiative}) in the nonminimal 
supersymmetric model and compared the predictions with those of
the minimal supersymmetric standard model~\cite{Basu:2007ys}.
This study was carried out for unpolarized electron and positron
beams. In this paper we continue this study and consider the
radiative process~(\ref{radiative}) in the nonminimal 
supersymmetric standard model with polarized  electron and positron
beams to understand in detail  if the signal can be enhanced by the 
use of polarized beams.
Furthermore, the SM background photons
from radiative neutrino production process~(\ref{radiativenu})
with beam polarizations will have to be taken into account
for a proper analysis of the radiative neutralino production 
process~(\ref{radiative}). Beam polarizations could enhance the 
signal photons for the process~(\ref{radiative}) for NMSSM 
and reduce those from the SM background at the same time, 
which could lead to the enhancement of the statistics.
We will also consider supersymmetric background photons from
radiative sneutrino production process~(\ref{radiativesnu}) 
with polarized beams.
This is important if  sneutrino
production is kinematically accessible and if the sneutrino decay is
invisible.  
We will compare and contrast the results obtained for NMSSM 
with those for the minimal supersymmetric standard model
with polarized beams.
This will include the signal for the radiative neutralino
process, and the  dependence of the cross sections on the parameters 
of the neutralino sector. This comparison will allow us
to assess the feasibility of observing the radiative neutralino 
process for the most popular low energy supersymmetric models
at a $e^+ e^-$ collider.   

The plan of the paper is as follows. In Sec.~\ref{sec: signal}, 
we calculate the  cross section for the signal process (\ref{radiative}) 
in the nonminimal supersymmetric standard model, and compare
it with the corresponding cross section in the minimal
supersymmtric standard model, for unpolarized and polarized
electron and positron beams.  In order to calculate the 
cross section in the NMSSM, we fix  the parameter space that we
use in our calculations. This is done by using various
theoretical and experimental constraints on the parameter space of 
NMSSM. In particular, we constrain the values of the trilinear 
superpotential parameters $\lambda$ and $\kappa$ which enter the 
neutralino mass matrix of the NMSSM.  We also describe the phase space
for the signal process as well as the cuts on outgoing photon 
angle and energy that we use to regularize the infrared and collinear 
divergences in the tree level cross section. We then analyze numerically
the dependence of the cross section on the parameters of the neutralino 
sector, and on the selectron masses, for unpolarized and polarized 
beams. Here we also calculate the photon energy distribution
for the radiative production of the second lightest neutralino
in the NMSSM and compare it with the corresponding distribution
for the lightest neutralino for unpolarized and polarized
beams, respectively.

In Sec.~\ref{sec:backgrounds} we analyze the backgrounds to the radiative
neutralino  process (\ref{radiative}) with  polarized beams. This includes the
background from SM process (\ref{radiativenu}),
as well as from the supersymmetric process (\ref{radiativesnu})
for the case of polarized beams.
In Sec.~\ref{sec:pol} we study in detail the beam polarization
dependence of the cross section for the radiative neutralino 
production process  as well as for the backgrounds processes.
Here we also consider the statistical significance for measuring
the excess of photons from radiative neutralino production over the 
backgrounds, and calculate this quantity for NMSSM, and compare it 
with the corresponding results in MSSM. We summarize our 
results and conclusions in Sec.~\ref{sec:conclusions}. 
Our notations and results on the neutralino mass matrices 
and couplings are summarized 
in  Appendix~\ref{appendix: superpotential and couplings}.

\section{Radiative Neutralino Production}
\label{sec: signal}
\begin{figure}[h!]
{%
\unitlength=1.0pt
\includegraphics{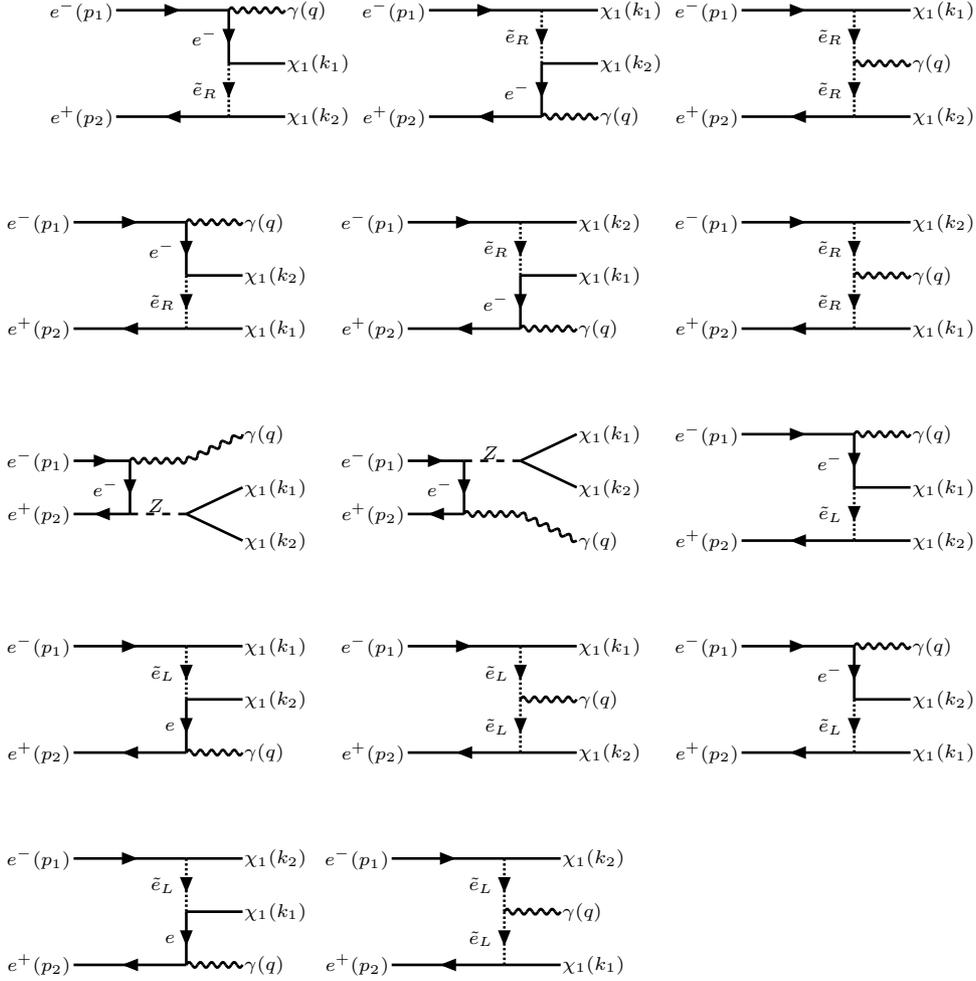}}
\caption{Feynman diagrams for the radiative production of
lightest neutralinos in the process $e^+e^- \to
 \tilde \chi{_1}^0 \tilde \chi{_1}^0 \gamma $.}
\label{fig:radneutralino}
\end{figure}

\begin{figure}[ht!]
\setlength{\unitlength}{1cm}
{\scalebox{1.3}{\includegraphics{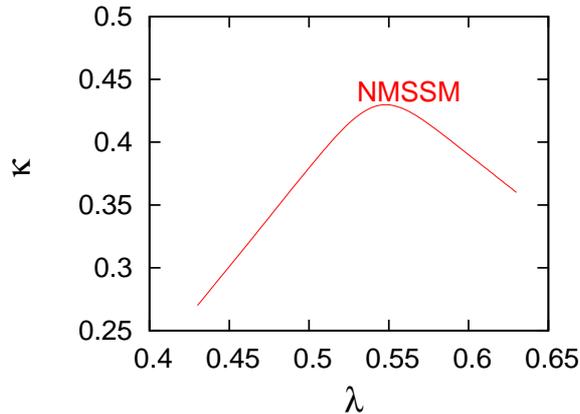}}}
\hspace{1mm}
\caption{Plot of $\lambda$ versus $\kappa$ for the set of input parameters
in  Table~\ref{parNMSSM}.}
\label{fig:lambda_kappa}
\end{figure}

\begin{figure}[ht!]
\setlength{\unitlength}{0.05cm}
\subfigure[
\label{fig:muM2N}]
{\scalebox{1.3}{\includegraphics{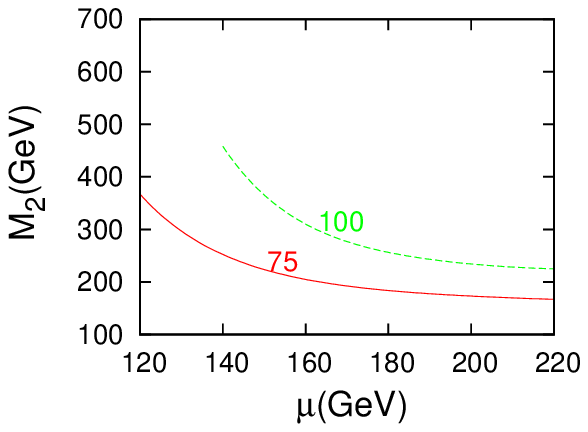}}}
\hspace{1mm}
\subfigure[
\label{fig:muM2M}]
{\scalebox{1.3}{\includegraphics{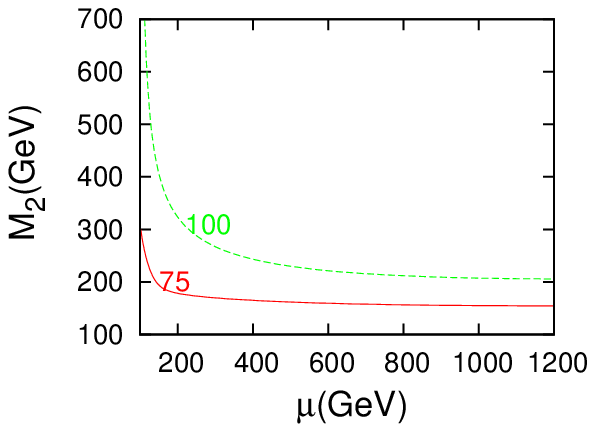}}}
\hspace{1mm}
\caption{(a) Contour plots  of constant lightest  neutralino
mass $m_{\chi^0_1}$ in
$\mu$ -$ M_2$ plane for NMSSM;~(b) for MSSM.}
\label{fig:muM2}
\end{figure}

The Feynman diagrams for the radiative process 
\begin{eqnarray}
e^-(p_1) + e^+(p_2) \rightarrow \tilde{\chi}_1^0(k_1) + \tilde{\chi}_1^0(k_2)
+ \gamma(q),
\label{radiative1}
\end{eqnarray}
are shown in Fig.~\ref{fig:radneutralino}, where the symbols in the brackets 
denote the four momenta of the respective particles. 
In NMSSM, and in MSSM, this process 
proceeds at the tree level via $t$- and $u$-channel exchange of
right and left selectrons $\tilde e_{R,L}$, and via $Z$  boson exchange
in the $s$ channel. 
In order to calculate the cross section for the
radiative production of neutralinos
we need to compute the couplings of the neutralinos to electrons, to
the scalar partners of electrons, the selectrons, and to $Z^0$ bosons.  
We summarize these couplings for  MSSM and 
NMSSM~\cite{ Haber:1984rc, Bartl:1989ms}
in Appendix~\ref{appendix: superpotential and couplings}.
As can be seen in  Appendix~\ref{appendix: superpotential and couplings},
the couplings of the lightest neutralino
are determined by the corresponding elements of the neutralino mixing
matrix~($N_{ij}$  or $N'_{ij}$). For numerical calculation of the radiative
neutralino cross section in the MSSM, we have chosen to work with
the parameters  in the Snowmass Points and Slopes~(SPS~1a) 
scenario~\cite{Allanach:2002nj}.
The parameters of the SPS~1a scenario are
summarized in Table~\ref{parMSSM}.
However, since in the SPS~1a scenario the value
of the parameters $\mu$ and $M_2$ are fixed, we shall use a different set of
parameters to study the dependence of the neutralino mass and the radiative
neutralino production cross section on $\mu$ and $M_2,$ and on the
selectron masses.  This set of parameters is shown in Table~\ref{parMSSMEWSB}.
We shall call this set of parameters as the MSSM electroweak symmetry
breaking scenario~(EWSB)~\cite{Pukhov:2004ca}. As in  
paper I, for the NMSSM we use a set of parameters that is obtained
by imposing theoretical and experimental constraints on the parameter 
space of the NMSSM. The parameters that enter the neutralino mass 
matrix of the NMSSM are, apart from
$M_1$ and $M_2$, $\tan\beta, \ \mu~(\equiv \lambda <S> = \lambda x), \  
\lambda$ and $\kappa$. For $M_1, M_2$ and $M_3$ we use the values which are 
consistent with the usual GUT relation 
$M_1/\alpha_1 = M_2/\alpha_2 = M_3/\alpha_3.$
We note that for the MSSM in  SPS~1a scenario, the value of the parameter 
$\tan\beta = 10 $. In order to remain close to the SPS~1a scenario of MSSM, 
we have chosen for our numerical calculations in NMSSM 
values of $\tan\beta = 10,$  whereas
the rest of the parameters  are chosen in such a way that the 
lightest Higgs boson mass, 
the lightest neutralino mass and the lightest chargino mass 
satisfy the present experimental lower limits. We have also imposed on 
the parameter space of 
NMSSM, the theoretical constraint that there is no charge and
color breaking global minimum of the scalar potential, and that a
Landau pole does not develop below the grand unified 
scale~($M_{GUT} \approx 10^{16}$ GeV). 
The consequence  of imposing
these constraints on the parameter space of NMSSM, and the resulting  masses
for various particles for a particular choice of input parameters
is summarized in Table~\ref{parNMSSM}.
\begin{table}[h!]
\renewcommand{\arraystretch}{1.0}
\caption{Input parameters and resulting  masses for various states in the 
MSSM SPS~1a scenario.}
\begin{center}
        \begin{tabular}{|c|c|c|c|c|}
\hline\hline
$\tan\beta=10$ & $Q= 100$~GeV & $m_{1/2}= 250$~GeV & $m_0=100$~GeV &$A_0 =-100$ ~GeV\\
$m_{\chi^0_{1}}=97$~GeV & 
$m_{\chi^\pm_{1}}=180$~GeV & 
$m_{\tilde e_{R}}=136$~GeV & 
$m_{\tilde\nu_e}=185$ GeV &
$m_h=110$ GeV\\
$m_{\chi^0_{2}}=180$~GeV & 
$m_{\chi^\pm_{2}}=379$~GeV & 
$m_{\tilde e_{L}}=195$~GeV &
$m_H =396$ ~GeV &
$m_A =395$ ~GeV\\
\hline\hline
\end{tabular}
\end{center}
\renewcommand{\arraystretch}{1.0}
\label{parMSSM}
\end{table}
\begin{table}[h!]
\renewcommand{\arraystretch}{1.0}
\caption{Input parameters and resulting masses of various states in  
MSSM~EWSB scenario.}
\begin{center}
        \begin{tabular}{|c|c|c|c|c|}
\hline\hline
$\tan\beta=10$ & $\mu= 149$~GeV & $M_1= 150$~GeV & $M_2=300$~GeV &$M_3 =1050$ ~GeV\\
$M_A=242$~GeV & $A_t= 3000$~GeV & $A_b= 3000$~GeV & $A_{\tau}=1000$~GeV & \\
$m_{\chi^0_{1}}=108$~GeV & 
$m_{\chi^\pm_{1}}=135$~GeV & 
$m_{\tilde e_{R}}=137$~GeV & 
$m_{\tilde\nu_e}=187$ GeV &
$m_h=118$ GeV \\
$m_{\chi^0_{2}}=-160$~GeV & 
$m_{\chi^\pm_{2}}=328$~GeV & 
$m_{\tilde e_{L}}=197$~GeV &
$m_H =243$ ~GeV &
$m_A =242$ ~GeV \\
\hline\hline
\end{tabular}
\end{center}
\renewcommand{\arraystretch}{1.0}
\label{parMSSMEWSB}
\end{table}
\begin{table}[h!]
\renewcommand{\arraystretch}{1.0}
\caption{Input parameters and resuling masses of various states in  NMSSM.}
\begin{center}
        \begin{tabular}{|c|c|c|c|c|}
\hline\hline
$\tan\beta=10$ & $\mu= 149$~GeV & $M_1= 150$~GeV & $M_2=300$~GeV &$M_3 =1050$ ~GeV\\
$M_A=242$~GeV & $A_t= 3000$~GeV & $A_b= 3000$~GeV & $A_{\tau}=1000$~GeV &$\lambda=0.54$\\
$\kappa=0.45$ & $A_{\lambda}=880$ ~GeV & $A_{\kappa}=10$ ~GeV & $m_{\tilde e_{R}}=137$~GeV & $m_{\tilde e_{L}}=197$~GeV \\
$m_{\chi^0_{1}}=94$~GeV & 
$m_{\chi^\pm_{1}}=135$~GeV & 
$m_{\tilde\nu_e}=187$ ~GeV &
$m_h=122$ ~GeV & \\
$m_{\chi^0_{2}}=-160$~GeV & 
$m_{\chi^\pm_{2}}=328$~GeV & 
$m_{H_2} =242$ ~GeV &
$m_{H_3} =1313$ ~GeV & \\
\hline\hline
\end{tabular}
\end{center}
\renewcommand{\arraystretch}{1.0}
\label{parNMSSM}
\end{table}
Since the neutralino mass matrix depends on the parameters $\lambda$ and
$\kappa$, it is useful to study the possible values of these parameters, 
with all other parameters fixed, which satisfy the phenomenological
and theoretical constraints discussed above. In Fig.~\ref{fig:lambda_kappa}
we show a  plot of  $\lambda$ versus $\kappa$, with all other input
parameters fixed as in Table~\ref{parNMSSM}, and with the
lightest neutralino, the lightest Higgs boson, 
and the lightest chargino masses as in  Table~\ref{parNMSSM} with a
variation of
less than $5\%$.  Fig.~\ref{fig:lambda_kappa} shows the range of  
$\lambda$ and $\kappa$ values that are consistent with all the constraints 
discussed above for the set of input parameters in Table~\ref{parNMSSM}.
We note that for the set of input values in Table~\ref{parNMSSM}, values of
$\lambda \lsim 0.4$, with  $ \kappa \lsim 0.22$,
lead to an unphysical global minimum. On the other hand,
values of $\lambda \gsim 0.57$, with  $ \kappa \gsim 0.45$, 
lead to a Landau pole below the GUT scale.
Thus, the allowed values of $\lambda$ and $\kappa$, for the given set of 
input parameters, and for the fixed masses of lightest neutralino, 
the lightest Higgs boson, and the lightest chargino, 
as in Table~\ref{parNMSSM},
lie in a narrow range   $0.4 \lsim \lambda \lsim 0.57$ 
for $ 0.22 \lsim \kappa \lsim 0.45$.
For definiteness, we have chosen to work with the values of
$\lambda = 0.54$ and $\kappa = 0.45$ in this paper. These values correspond
to the peak in the $\lambda$ versus $\kappa$ plot in 
Fig.~\ref{fig:lambda_kappa}. 
For the  parameters of Table~ \ref{parNMSSM}, the composition of the 
lightest neutralino in NMSSM is given by
\begin{eqnarray}
N'_{1j} & = & (0.48,~ -0.23,~ 0.57,~ -0.55,~  0.30).
\label{nmssmcomp}
\end{eqnarray}
{}From the composition (\ref{nmssmcomp}), we see that the lightest neutralino
has a sizable singlet component, thereby changing the  neutralino
phenomenology in the NMSSM as compared to MSSM. 
For comparison, we also show the particle content
of the lightest neutralino in MSSM
\begin{eqnarray}
N_{1j} & = & (0.6,~ -0.21,~ 0.61,~ -0.47),
\label{mssmcomp}
\end{eqnarray}
for the parameter set in Table~\ref{parMSSMEWSB}. In  Fig.~\ref{fig:muM2N} 
we have plotted the constant contour plots  for the mass
of lightest neutralino  in NMSSM in the $\mu$ - $M_2$ plane. We emphasize
that the choice of $\mu$ and  $M_2$ values in this plot  have been taken to be 
consistent with phenomenological and theoretical constraints as
described above. 
For comparison, we have also plotted the corresponding contour 
plots  for MSSM in
Fig.~\ref{fig:muM2M} with parameters as in Table ~\ref{parMSSMEWSB}. 

\begin{figure}[h!]
\setlength{\unitlength}{0.05cm}
\subfigure[
\label{fig:neutralinodiffun}]{\scalebox{1.3}{\includegraphics{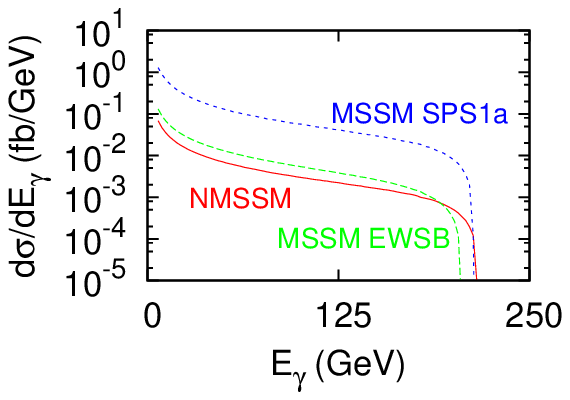}}}
\hspace{1mm}
\subfigure[
\label{fig:neutralinodiffp}]{\scalebox{1.3}{\includegraphics{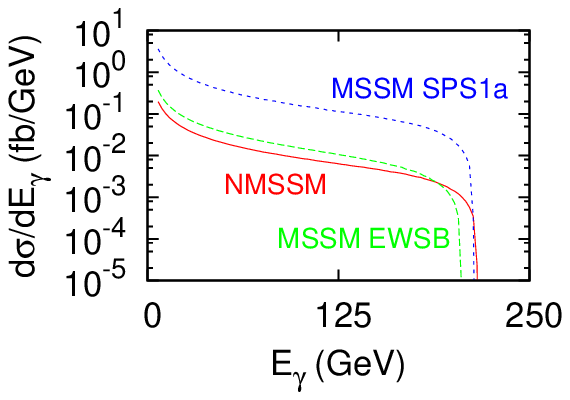}}}
\caption{ (a) Photon energy
          distribution  {$\displaystyle \frac{d\sigma}{d E_\gamma}$}
          for the radiative neutralino production for
          NMSSM~(red solid line),  for MSSM EWSB~(green dashed line) and 
          MSSM SPS~ 1a~(blue dashed line) at $\sqrt{s} =
          500$ GeV with ($P_{e^-}$,  $P_{e^+}$) =(0, 0);~(b) with 
          ($P_{e^-}$,  $P_{e^+}$) =(0.8, - 0.6).}
\label{fig:neutralinodiff}
\end{figure}

\subsection{Cross Section for the Signal Process}
In NMSSM, and in MSSM, the process  (~\ref{radiative1}) proceeds at the tree 
level via $t$- and $u$-channel exchange of right and left selectrons
$\tilde e_{R,L},$  and via $Z$ boson exchange in the $s$ channel.
The photon is radiated off the incoming beams or the exchanged selectrons. 
The corresponding Feynman diagrams are shown in Fig.~\ref{fig:radneutralino}.
The differential cross section for (~\ref{radiative1}) can be written 
as ~\cite{Grassie:1983kq, Eidelman:2004wy}
\begin{eqnarray}
\label{sec:results}
d \sigma &=& \frac{1}{2} \frac{(2\pi)^4}{2 s}
\prod_f \frac{d^3 \mathbf{p}_f}{(2\pi)^3 2E_f}\delta^{(4)}(p_1 +
p_2 - k_1 - k_2 - q)|\M|^2,
\label{phasespace}
\end{eqnarray}
where $\mathbf{p}_f$ and $E_f$ denote the final three-momenta
$\mathbf{k}_1$, $\mathbf{k}_2$, $\mathbf{q}$
and the final energies
$E_{\chi_1}$, $E_{\chi_2}$, and $E_\gamma$
of the neutralinos and the photon, respectively.
The squared matrix element $|\M|^2$ in (~\ref{phasespace})
can be written as~\cite{Grassie:1983kq}
\begin{eqnarray}
|\M|^2 & = & \sum_{i \leq j} T_{ij}, \label{squaredmatrix}
\end{eqnarray}
where $T_{ij}$ are squared amplitudes corresponding to the Feynman diagrams
in  Fig.~\ref{fig:radneutralino}. A sum over the
spins of the outgoing neutralinos, as well as a sum over the polarizations
of the outgoing photon is included in $T_{ij}$.
We have included the longitudinal beam polarizations of electrons,
$P_{e^-}$, and positrons, $P_{e^+}$, with  $-1 \le P_{e^\pm} \le +1$,
while calculating the cross section for the
the radiative neutralino production process.
The phase space in (~\ref{phasespace}) is described in~\cite{Grassie:1983kq}.

\begin{figure}[h!]
\setlength{\unitlength}{0.05cm}
\subfigure[
\label{fig:neutralinototalun}]{\scalebox{1.3}{\includegraphics{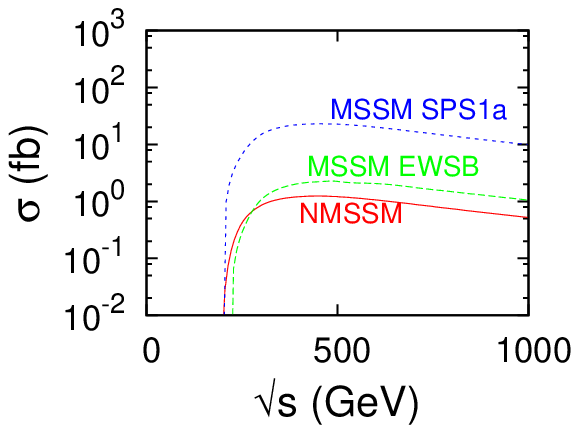}}}
\hspace{1mm}
\subfigure[
\label{fig:neutralinototalp}]{\scalebox{1.3}{\includegraphics{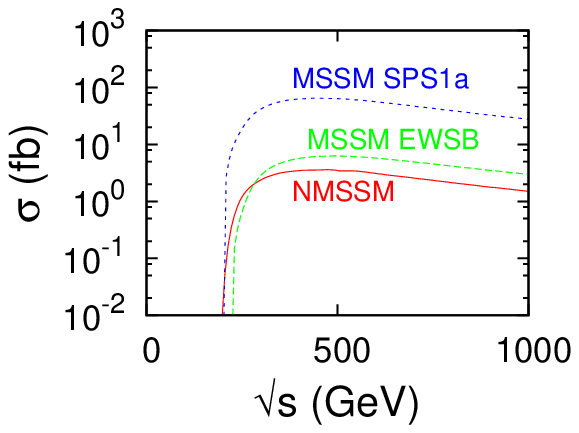}}}
\caption{(a) Total energy $\sqrt{s}$ dependence of
          the cross sections $\sigma$ 
         for radiative neutralino production $e^+e^- \to \tilde\chi^0_1
         \tilde\chi^0_1\gamma$ for NMSSM~(red solid line) and for MSSM  EWSB 
         scenario~(green dashed line)and MSSM SPS~1a~(blue dashed line) with 
($P_{e^-}$, $P_{e^+}$)=(0, 0),~(b) with ($P_{e^-}$,  $P_{e^+}$) =(0.8, - 0.6).}
\label{fig:neutralinototal}
\end{figure}

\subsubsection{Numerical Results}
We have calculated the squared amplitudes and the tree-level cross section 
for radiative neutralino production
~(\ref{radiative}), and the  background from radiative neutrino and 
sneutrino production,
~(\ref{radiativenu}) and~(\ref{radiativesnu}), with polarized
electron and positron beams, using the program CALCHEP~\cite{Pukhov:2004ca}. 
We note that when integrating the squared amplitude for the
radiative neutralino production, the $s - t$ interference terms
cancel the $s - u$ interference terms due to a symmetry in these channels,
due to the Majorana properties of the neutralinos~\cite{Choi:1999bs}.
The tree level cross sections have infrared and collinear divergences, 
which need to be regularized~\cite{ Grassie:1983kq}. 
To do this we define the fraction of the
beam energy carried by the photon as $x = E_{\gamma}/E_{\rm beam},$
where  $ \sqrt {s} = 2E_{\rm beam}$ is the center of mass energy, and 
$E_{\gamma}$ is the energy carried away by the photon. 
We then impose the following cuts on $x$,  and on the scattering 
angle $\theta_{\gamma}$ of the 
photon~\cite{Dreiner:2006sb}:
\begin{eqnarray}
0.02 \le x \le  1-\frac{m_{\chi_1^0}^2}{E_{\rm beam}^2}, \label{cut1} \\
\nonumber \\
-0.99 \le \cos\theta_\gamma \le 0.99.
\label{cut2}
\end{eqnarray}
The lower limit on  $x$ in (\ref{cut1}) corresponds to a photon energy 
$E_\gamma= 5$\, GeV for the center of mass energy $\sqrt{s}=500$\, GeV. The
upper limit of  $(1-m_{\chi_1^0}^2/E_{\rm beam}^2)$ on  $x$  
corresponds to the maximum energy that a photon can carry
in radiative neutralino production. 

In order to implement the cuts on the photon energy in the calculation
of the cross sections, we have taken the  mass of the lightest
neutralino in NMSSM to be $m_{\chi_1^0}=94$~GeV for the parameter 
set shown in Table~\ref{parNMSSM}.
For MSSM SPS1a, we take $m_{\chi_1^0}=97$~GeV and for MSSM EWSB  $m_{\chi_1^0}=108$~GeV.

We note that for values of  $\sqrt{s}=500$ GeV, and 
for $m_{\chi_1^0} \geq 94$ GeV,
this cut reduces a substantial amount of the on-shell $Z$ boson 
contribution to radiative neutrino production process.

\subsubsection{Photon Energy~($E_\gamma$) Distribution and Total Beam
Energy~($\sqrt {s}$) Dependence}
Using the procedure described above, 
we have calculated  the energy distribution of the photons from
radiative neutralino  
production in  NMSSM,  in MSSM SPS~1a, and in MSSM EWSB
for both unpolarized 
and polarized electron and positron beams, respectively.  
These are shown in
Fig.~\ref{fig:neutralinodiff},  where  we  compare  the
energy distribution of the photons in  these models.
In Fig.~\ref{fig:neutralinototal} we  show the total beam energy $\sqrt s$ 
dependence of the cross sections  for NMSSM,
and for MSSM ~EWSB and MSSM SPS~1a,  respectively.   
We note that the photon energy distribution and the total cross section
for radiative neutralino production in NMSSM and in MSSM EWSB are
very close to each other, and are smaller as compared to what one obtains
in MSSM SPS 1a scenario.
\subsubsection{Dependence on  $\mu$ and $M_2$}
\begin{figure}[h!]
\setlength{\unitlength}{0.05cm}
\subfigure[
\label{fig:neutralinomuun}]{\scalebox{1.3}{\includegraphics{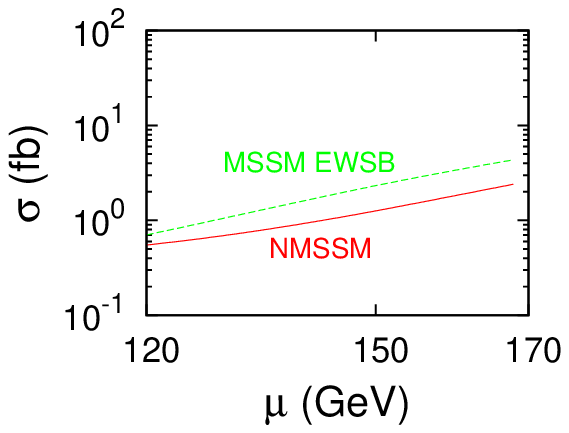}}}
\hspace{1mm}
\subfigure[
\label{fig:neutralinomup}]{\scalebox{1.3}{\includegraphics{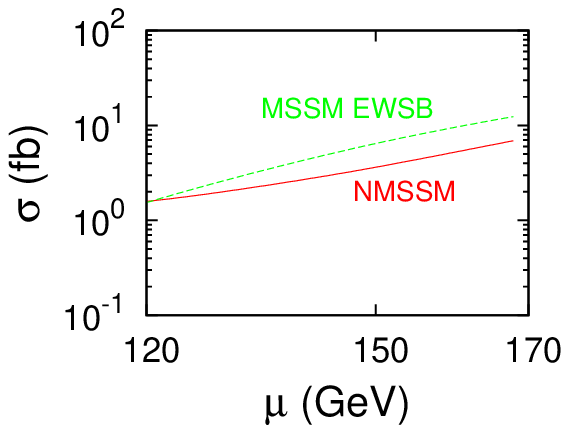}}}
\caption{ (a) Total cross-section $\sigma$ for  the radiative neutralino 
          production  versus $\mu$ for NMSSM~(red solid line) and for
          MSSM EWSB scenario~(green dashed) at $ \sqrt {s} = 
          500 $ GeV with ($P_{e^-}$,  $P_{e^+}$)=(0, 0);~(b) 
          with ($P_{e^-}$,  $P_{e^+}$) =(0.8, - 0.6).}
\label{fig:neutralinomu}
\end{figure}
\begin{figure}[h!]
\setlength{\unitlength}{0.05cm}
\subfigure[
\label{fig:neutralinoM2un}]{\scalebox{1.3}{\includegraphics{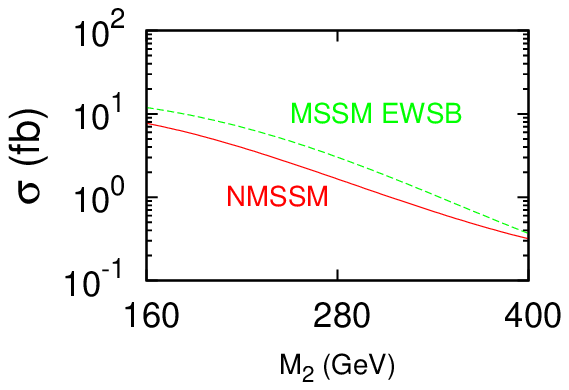}}}
\hspace{1mm}
\subfigure[
\label{fig:neutralinoM2p}]{\scalebox{1.3}{\includegraphics{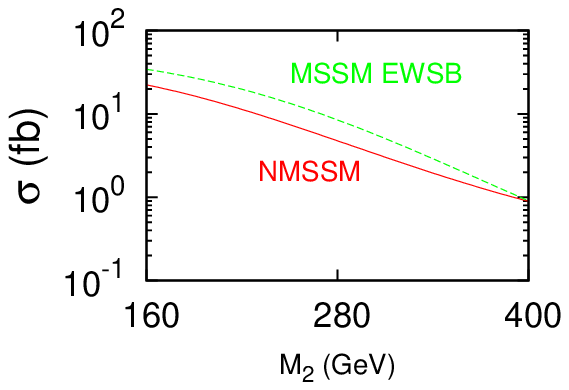}}}
\caption{(a) Total cross section $\sigma$ for  the radiative neutralino 
production versus $M_2$ for NMSSM~(red solid line) and for 
MSSM EWSB scenario~(green dashed) at $ \sqrt {s} = 500$ GeV 
with ($P_e^-$, $P_e^+$)=(0,  0);~(b) with ($P_{e^-}$,  i
$P_{e^+}$) =(0.8, - 0.6).}
\label{fig:neutralinoM2}
\end{figure}

Since the neutralino mass matrix, and hence the lightest neutralino mass, depends on
$\mu$ and $M_2$, it is important  to study the dependence of the radiative neutralino
cross section on these parameters.
In the nonminimal supersymmetric standard model, $\mu~(\equiv \lambda <S> = \lambda x)$ 
and $M_2$ are independent parameters. We have, therefore, studied the  cross section 
$\sigma$($e^+e^- \to \tilde\chi^0_1 \tilde\chi^0_1\gamma$) as a function of
$\mu$ and  $M_2$ independently. 
In Fig.~\ref{fig:neutralinomu}  we show the $\mu$ dependence
of the total cross section for the radiative production of neutralinos 
for  NMSSM as well as MSSM~EWSB. We recall that in the MSSM SPS~1a
scenario these parameters are fixed. As is seen from Fig.~\ref{fig:neutralinomu},
the total cross section increases with $\mu$. The plot  of total cross-section versus 
$\mu$ in Fig.~\ref{fig:neutralinomu} is plotted in the range $\mu \in [120,170]$~GeV in
NMSSM and in MSSM~EWSB. Note that the parameter values are chosen 
so as to avoid color and charge breaking minima, absence of Landau pole, and
the phenomenological constraints on different particle masses. 
Furthermore, in Fig.~\ref{fig:neutralinoM2}  we show the $M_2$ dependence of 
the total cross section for radiative neutralino production for NMSSM and 
MSSM~EWSB. The total cross-section decreases with increasing value of $M_2$. 
The graph of total cross-section versus $M_2$ in Fig.~\ref{fig:neutralinoM2} 
is plotted for the interval $M_2 \in [150,450]$~GeV in NMSSM and in MSSM~EWSB 
so as to satisfy the theoretical and phenomenological constraints described 
above. From Figs.~\ref{fig:neutralinomu} and \ref{fig:neutralinoM2} we note
that the cross section is significantly enhanced  when the electron 
and positron 
beams are polarized as compared to the case when the beams are unpolarized.

\subsubsection{Dependence on selectron masses}
\begin{figure}[h!]
\setlength{\unitlength}{1cm}
\subfigure[
\label{fig:neutralinoselLun}]{\scalebox{0.65}{\includegraphics{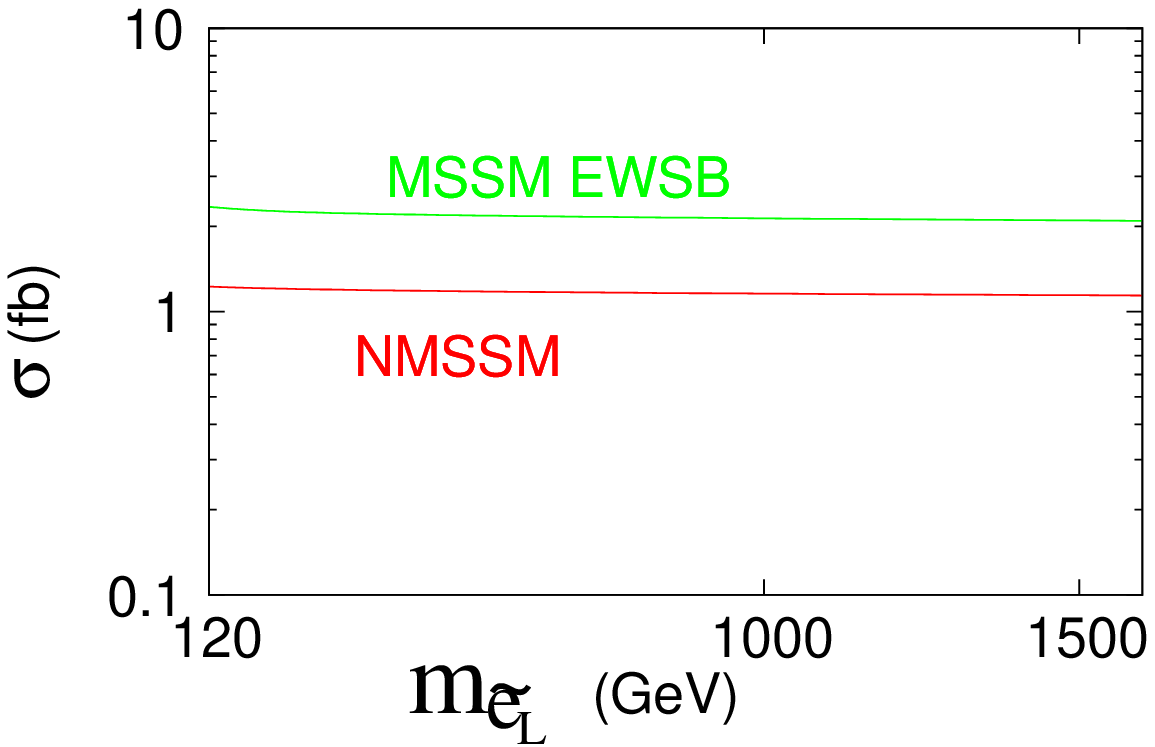}}}
\hspace{1mm}
\subfigure[
\label{fig:neutralinoselLp}]{\scalebox{0.65}{\includegraphics{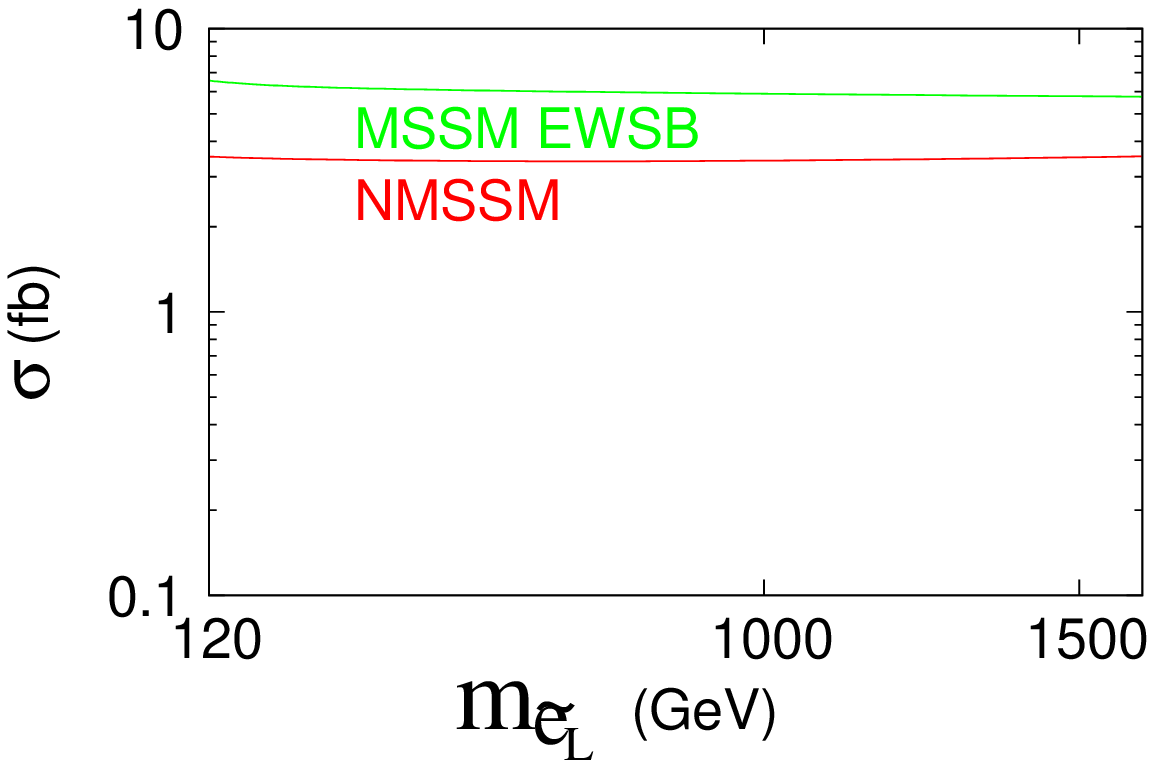}}}
\caption{ (a) Total cross section $\sigma$ for  the radiative neutralino 
          production  versus $ m_{\tilde e_L}$ for NMSSM~(red solid line) 
          and for MSSM in the  EWSB scenario~(green dashed) at 
          $ \sqrt {s} = 500$ GeV with ($P_{e^-}$,   $P_{e^+}$)=(0, 0),
          ~(b) with ($P_{e^-}$,  $P_{e^+}$) =(0.8, - 0.6).}
\label{fig:neutralinoselL}
\end{figure}

\begin{figure}[h!]
\setlength{\unitlength}{1cm}
\subfigure[
\label{fig:neutralinoselRun}]{\scalebox{0.65}{\includegraphics{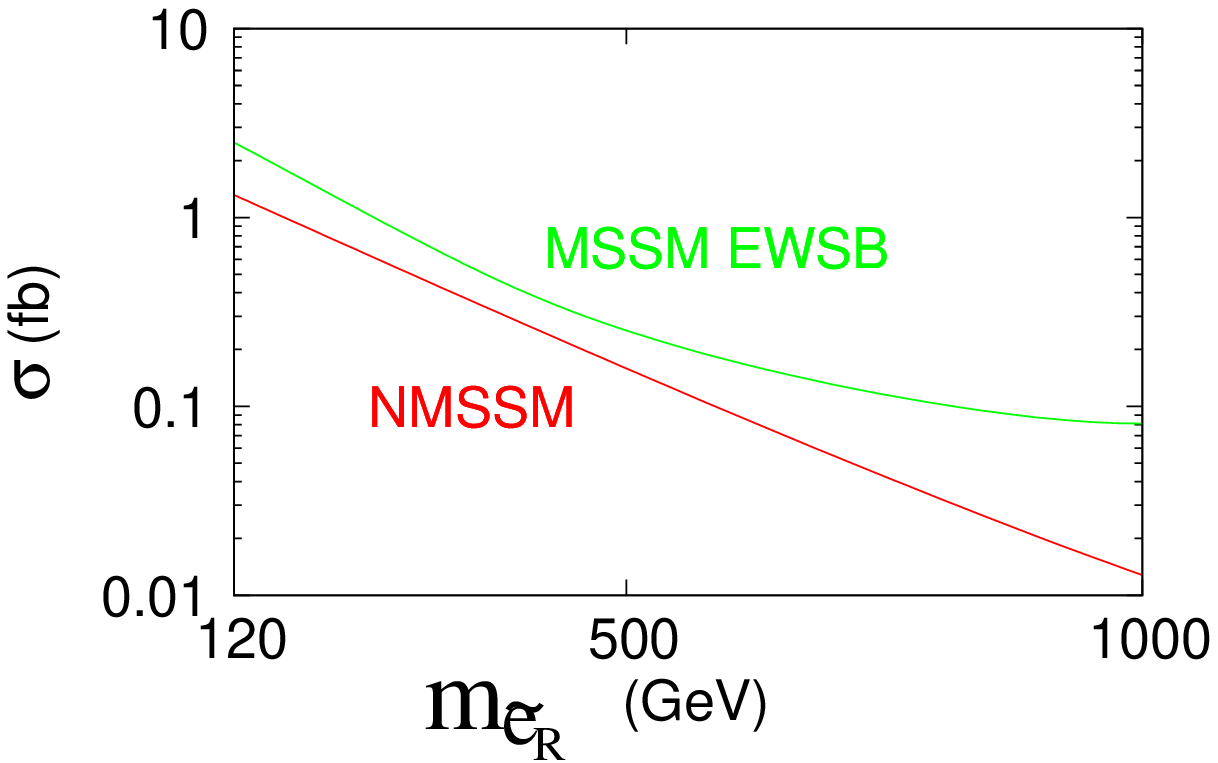}}}
\hspace{1mm}
\subfigure[
\label{fig:neutralinoselRp}]{\scalebox{0.65}{\includegraphics{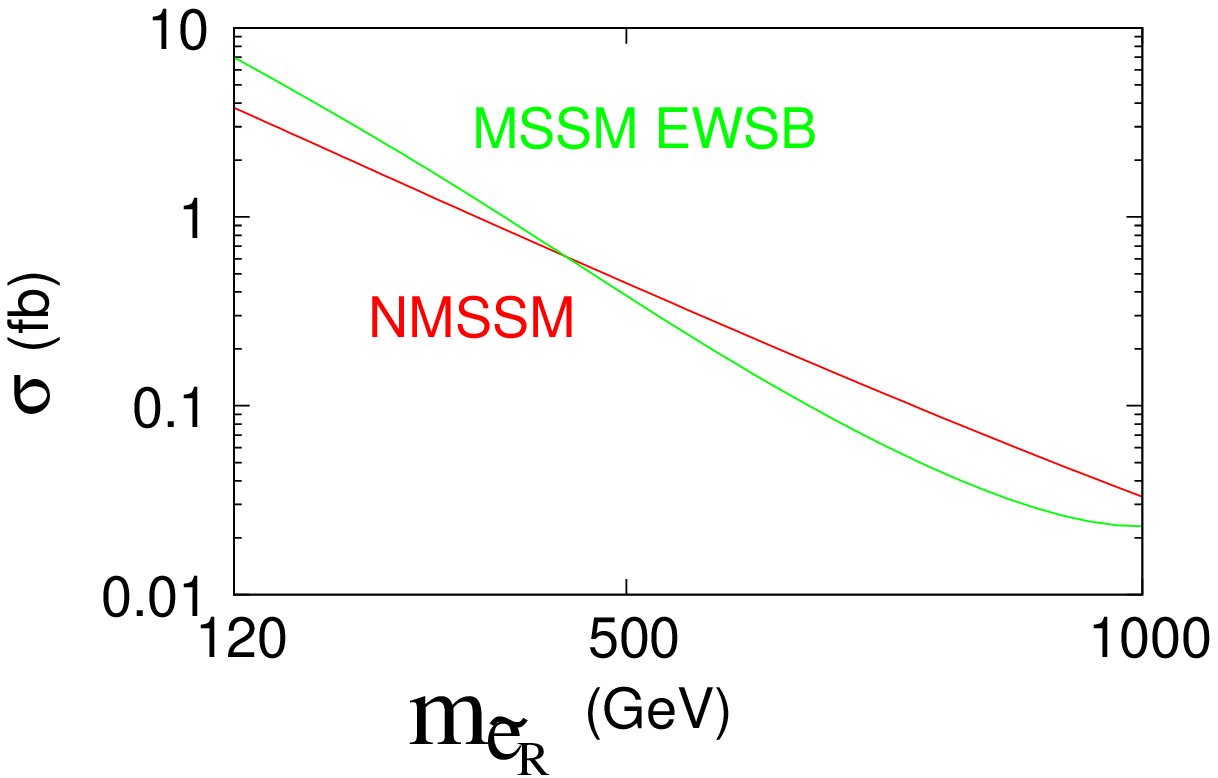}}}
\caption{ (a) Total cross section $\sigma$ for  the radiative neutralino
          production  versus $m_{\tilde e_R}$ for NMSSM~(red solid line) and for 
         MSSM in the EWSB scenario~(green dashed) at $ \sqrt {s} = 500$ GeV with 
($P_{e^-}$,   $P_e^+$)=(0, 0), ~(b) with ($P_{e^-}$,  $P_{e^+}$) =(0.8, - 0.6).}
\label{fig:neutralinoselR}
\end{figure}

The cross section for radiative neutralino production $\sigma(e^+e^-
\to\tilde\chi^0_1\tilde\chi^0_1\gamma)$ proceeds mainly via right and left
selectron $\tilde e_{R,L}$ exchange in the $t$ and $u$-channels.  In the NMSSM
and MSSM~EWSB,  the selectron masses are independent parameters. 
In Fig.~\ref{fig:neutralinoselL} and Fig.~\ref{fig:neutralinoselR} we  show the 
dependence   of total cross section of radiative neutralino production 
on the left and right selectron masses. The cross section is not very 
sensitive to the selectron masses for both models. Furthermore, the total 
neutralino production 
cross section is smaller in NMSSM as compared to MSSM~EWSB as
a function of left as well as right selectron masses.
  Again we note that the cross sections, as a function of selectron masses,
are larger in the case of polarized beams as compared to the 
unpolarized case.
\subsubsection{Photon energy~($E_\gamma$) distribution
for the production of the second lightest neutralino}
\begin{figure}[h!]
\setlength{\unitlength}{0.05cm}
\subfigure[
\label{fig:2ndneutralinoun}]{\scalebox{1.3}{\includegraphics{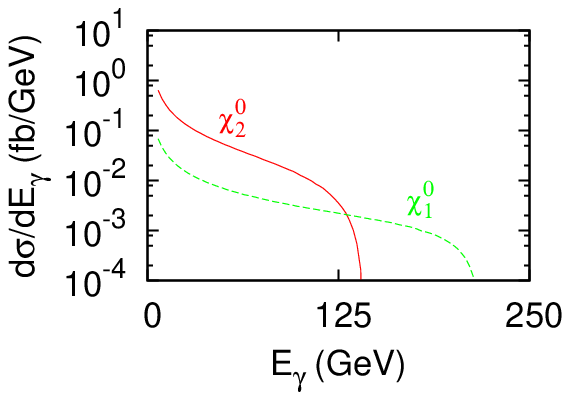}}}
\hspace{1mm}
\subfigure[
\label{fig:2ndneutralinop}]{\scalebox{1.3}{\includegraphics{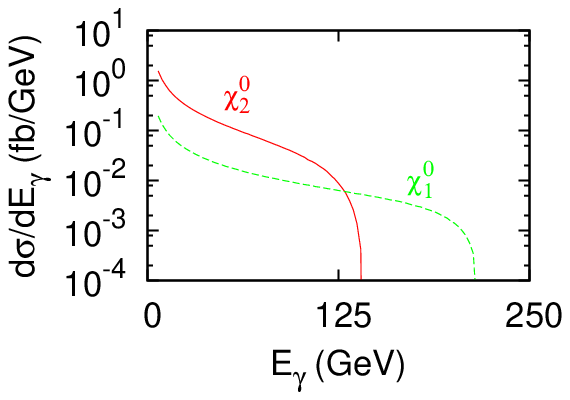}}}
\caption{ (a) Photon energy
          distribution  {$\displaystyle \frac{d\sigma}{d E_\gamma}$}
          for the radiative production of second lightest neutralino
          in  NMSSM~(red solid line),  and the lightest neutralino 
          for NMSSM~(green dashed line)  at $\sqrt{s} =
          500$ GeV with ($P_{e^-}$,  $P_{e^+}$) =(0, 0);~(b) with 
          ($P_{e^-}$,  $P_{e^+}$) =(0.8, - 0.6).}
\label{fig:2ndneutralino_1stneutralino}
\end{figure}
The cross section for the production of the lightest
neutralino in NMSSM is relatively small compared with the 
corresponding cross section for the lightest neutralino in the MSSM ~SPS1a. 
It may,  therefore,
be useful to consider the radiative production of the second lightest 
neutralino in the NMSSM. For the parameter set of Table~\ref{parNMSSM} the
composition of the second lightest neutralino in NMSSM is given by
\begin{eqnarray}
N'_{2j} & = & (0.87,~ 0.21,~ -0.22,~ 0.34,~ -0.19).
\label{nmssmcomp2}
\end{eqnarray}
We have calculated the photon energy distribution for the radiative 
production of the second lightest neutralino in NMSSM for the set of 
parameters shown in Table~\ref{parNMSSM}. This is shown in 
Fig.~\ref{fig:2ndneutralino_1stneutralino}. For comparison we have also 
shown the photon energy distribution for the radiative production of the 
lightest neutralino in NMSSM. We see that the cross section 
for the production of the second lightest neutralino is much 
smaller than  the cross section for the lightest neutralino at photon 
energy greater than 140 GeV. However, at lower photon energies the 
photon energy distribution for the second lightest neutralino
is significantly larger, both for unpolarized as well as polarized
beams.
\begin{figure}[h!]
\includegraphics{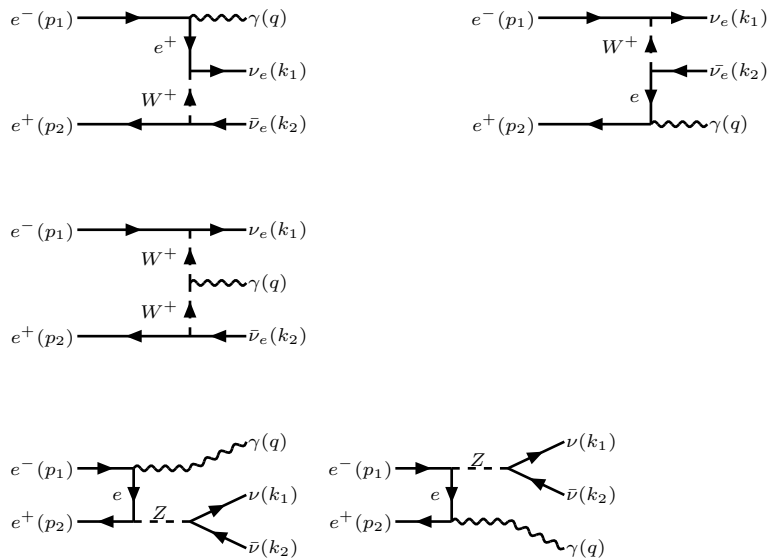}
\caption{Feynman diagrams contributing to the radiative neutrino process 
 $e^+e^- \rightarrow {\nu}{\bar{\nu}}\gamma.$}
\label{fig:radneutrino}
\end{figure}
\section {Background Processes}
\label{sec:backgrounds}
\subsection{The Neutrino Background}
\noindent
The major background to the radiative neutralino production~(~\ref{radiative1})
comes from the SM radiative neutrino production
process~\cite{Datta:1996ur,Gaemers:1978fe,Berends:1987zz,
Boudjema:1996qg,Montagna:1998ce}
\begin{equation}
e^+ +e^- \to \nu_\ell+\bar\nu_\ell+\gamma\,,\;\;\qquad \ell=e,\mu,\tau.
\label{radiative2}
\end{equation}
In this process $\nu_e$ are produced via
$t$-channel $W$ boson exchange, and $\nu_{e,\mu,\tau}$ via $s$-channel
$Z$ boson exchange.  We Feynman diagrams contributing to the process
(\ref{radiative2}) are shown in Fig.~\ref{fig:radneutrino}.  

\begin{figure}[t!]
\setlength{\unitlength}{0.05cm}
\subfigure[
\label{fig:neutrinodiffun}]{\scalebox{1.3}{\includegraphics{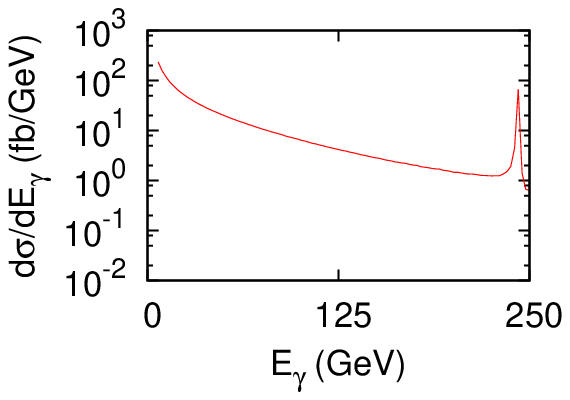}}}
\hspace{1mm}
\subfigure[
\label{fig:neutrinodiffp}]{\scalebox{1.3}{\includegraphics{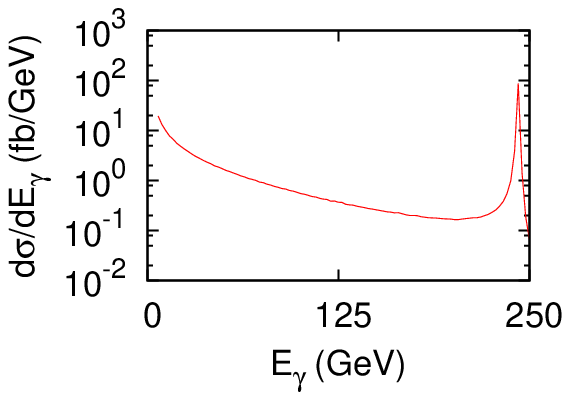}}}
\caption{ (a) The photon energy distribution $\frac{d\sigma}{d E_\gamma}$
        for the radiative  neutrino  process  
        $e^+e^- \to \nu\bar\nu\gamma$  at $\sqrt {s} = 500$ GeV  with 
($P_{e^-}$,   $P_e^+$)=(0, 0), ~(b) with ($P_{e^-}$,  $P_{e^+}$) =(0.8, - 0.6).}
\label{fig:neutrinodiff}
\end{figure}

\begin{figure}[t!]
\setlength{\unitlength}{0.05cm}
\subfigure[
\label{fig:neutrinototalun}]{\scalebox{1.3}{\includegraphics{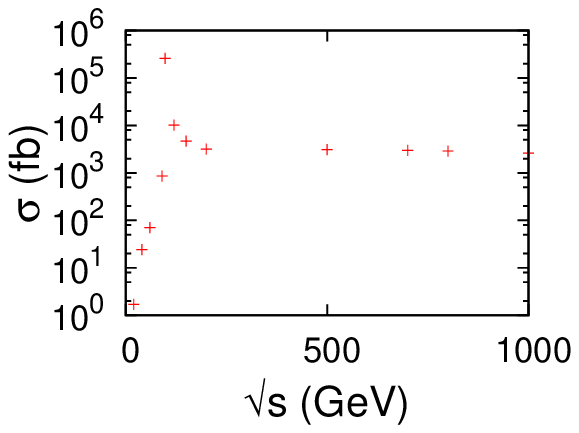}}}
\hspace{1mm}
\subfigure[
\label{fig:neutrinototalp}]{\scalebox{1.3}{\includegraphics{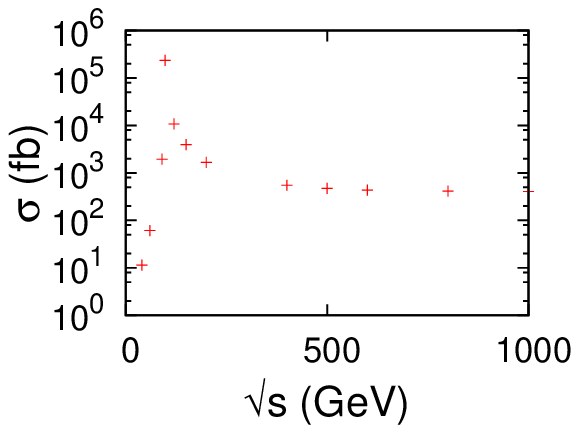}}}
\caption{ (a) Total energy $\sqrt{s}$ dependence of the cross sections $\sigma$ 
for radiative neutrino cross section $\sigma$($e^+e^- \to \nu\bar\nu\gamma$) with 
($P_{e^-}$, $P_{e^+}$)=(0, 0),~(b) with ($P_{e^-}$,  $P_{e^+}$) =(0.8, - 0.6).}
\label{fig:neutrinototal}
\end{figure}

The background photon energy distribution $\frac{d\sigma}{d E_\gamma}$ and 
$\sqrt s$ dependence of the cross
section $\sigma$ for radiative  neutrino  production $e^+e^- \to
\nu\bar\nu\gamma$ with polarized electron and positron beams
is the  same for both NMSSM and MSSM.  
As shown in Fig.~\ref{fig:neutrinodiff} the photon energy distribution from 
the radiative neutrino production peaks at
$E_\gamma= (s -m_Z^2)/(2\sqrt{s}) \approx 242$~GeV because of  
the radiative $Z$ production($\sqrt s > m_Z$). 
This  photon background from radiative neutrino production can be reduced  
by imposing an upper cut on the photon energy 
$x^{\rm max}=E_\gamma^{\rm max}/E_{\rm beam}=1-m_{\chi_1^0}^2
/E_{\rm beam}^2$ GeV in NMSSM, see Eq.~(\ref{cut1}), which is the kinematical
endpoint $E_\gamma^\mathrm{max}\approx 215$ GeV of the energy
distribution of the photon from radiative neutralino production
\begin{eqnarray}
 m_{\chi_1^0}^2 = \frac{1}{4}\left(s - 2\sqrt{s}E_\gamma^\mathrm{max}
\right). 
\end{eqnarray}
In order to achieve this, one would have to separate the signal
and background processes. This would be possible if the neutralino is
heavy enough, such that the endpoint is  removed from the
$Z^0$ peak of the background distribution. See also \cite{Dreiner:2006sb}.
In Fig.~\ref{fig:neutrinototal} we show the $\sqrt s$ dependence of the
total radiative neutrino cross section with unpolarized and polarized
electron and positron beams without imposing upper cut on the photon energy.
In Section~\ref{sec:pol} the upper cut on the photon
energy $E_\gamma^\mathrm{max}= 214.7$ GeV is used for the 
calculation of cross section for the radiative neutrino production.
\subsection{The Supersymmetric Background}
\begin{figure}[h!]
{%
\unitlength=1.0pt
\includegraphics{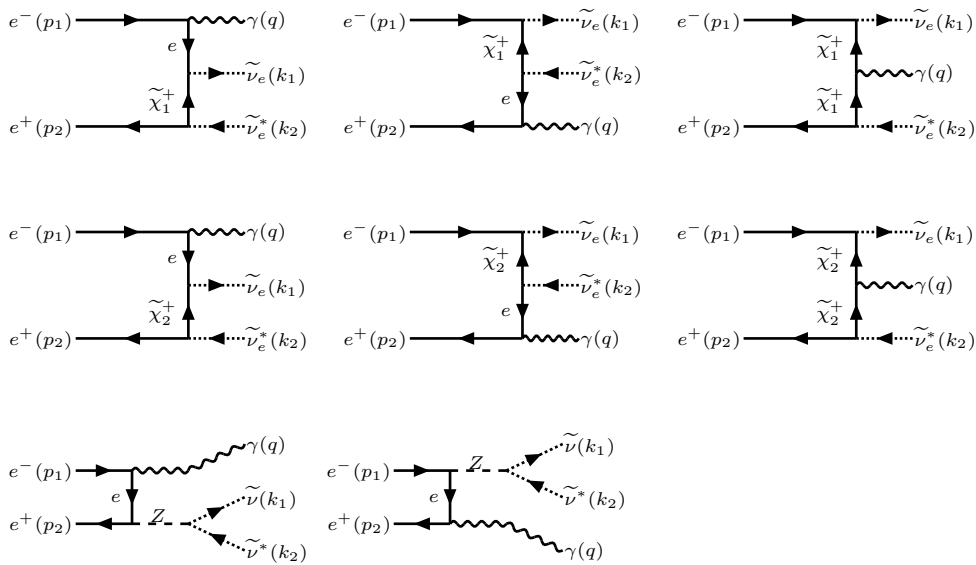}
}
\caption{Feynman diagrams contributing  to the radiative sneutrino production
process $e^+e^- \rightarrow \tilde{\nu}\tilde{\nu}^*\gamma.$}
\label{fig:radsneutrino}
\end{figure}

\begin{figure}[t!]
\setlength{\unitlength}{0.05cm}
\subfigure[
\label{fig:sneutrinodiffun}]{\scalebox{1.3}{\includegraphics{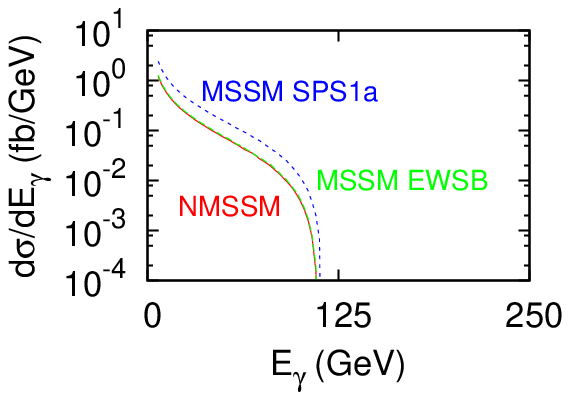}}}
\hspace{1mm}
\subfigure[
\label{fig:sneutrinodiffp}]{\scalebox{1.3}{\includegraphics{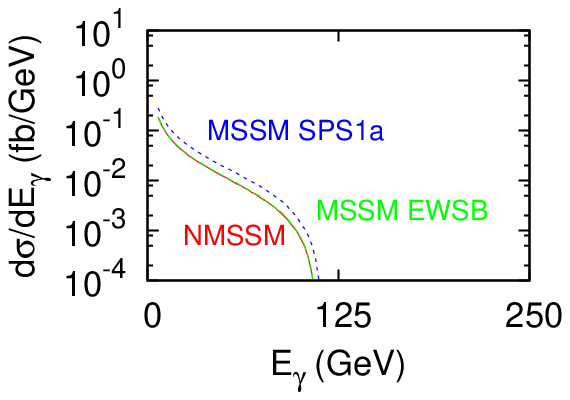}}}
\caption{ (a)The photon energy distribution $\frac{d\sigma}{d E_\gamma}$
        for the radiative  sneutrino  production 
         $e^+e^- \to \tilde\nu\tilde\nu^\ast\gamma$  
         at $\sqrt{s} = 500$ GeV with 
           ($P_{e^-}$,   $P_e^+$)=(0, 0), ~(b) with ($P_{e^-}$,  $P_{e^+}$) =(0.8, - 0.6).}
\label{fig:sneutrinodiff} 
\end{figure}

\begin{figure}[h!]
\setlength{\unitlength}{0.05cm}
\subfigure[
\label{fig:sneutrinototalun}]{\scalebox{1.3}{\includegraphics{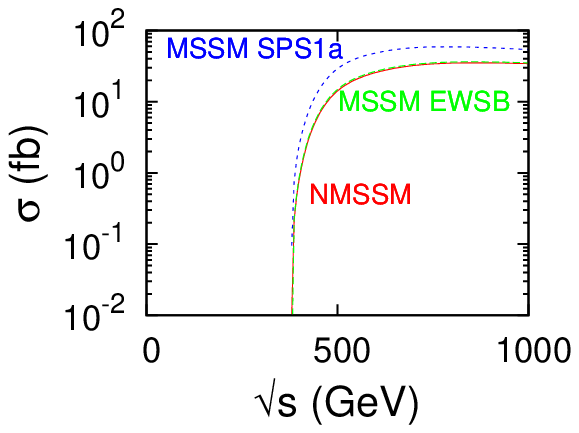}}}
\hspace{1mm}
\subfigure[
\label{fig:sneutrinototalp}]{\scalebox{1.3}{\includegraphics{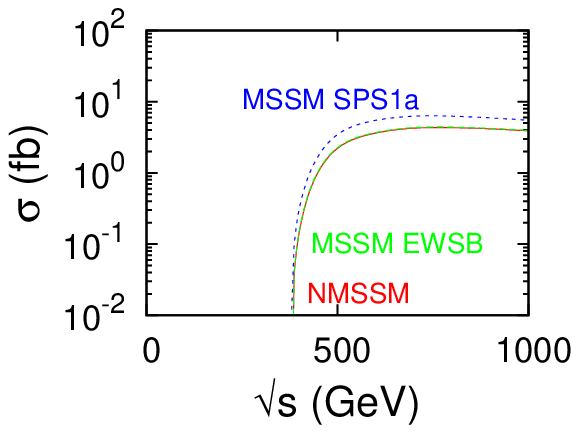}}}
\caption{ (a)Total energy $\sqrt {s}$
  dependence of the radiative  sneutrino production cross
  section $\sigma$($e^+e^- \to \tilde\nu\tilde\nu^\ast\gamma$)with 
($P_{e^-}$, $P_{e^+}$)=(0, 0),~(b) with ($P_{e^-}$,  $P_{e^+}$) =(0.8, - 0.6).}
\label{fig:sneutrinototal}
\end{figure}

Apart from the SM background coming from (\ref{radiative2}), the radiative
neutralino production  (\ref{radiative1}) has a background coming from the
supersymmetric sneutrino production process~\cite{Datta:1996ur, Franke:1994ph}
\begin{equation}
e^+ +e^- \to \tilde\nu_\ell+\tilde\nu^\ast_\ell+\gamma\,, 
\;\qquad \ell=e,\mu,\tau\,.
\label{radiative3}
\end{equation}
The lowest order Feynman diagrams contributing to the process 
(\ref{radiative3}) are shown in Fig.~\ref{fig:radsneutrino}.
This process receives $t$-channel contributions via
virtual charginos for $\tilde\nu_e\tilde\nu_e^\ast $-production, as
well as $s$-channel contributions from $Z$ boson exchange for
$\tilde\nu_{e, \mu,\tau}\tilde\nu_{e, \mu,\tau}^\ast $-production. 
In Fig.~\ref{fig:sneutrinodiff}, we show the photon energy distribution 
$\frac{d\sigma}{d E_\gamma}$ for radiative  sneutrino  production
$e^+e^- \to \tilde\nu\tilde\nu^\ast\gamma$ at $\sqrt{s} = 500$ GeV
with unpolarized and longitudinally polarized electron and positron beams.
The corresponding total cross section for the radiative sneutrino production
is shown in Fig.~\ref{fig:sneutrinototal}.


Radiative sneutrino production (~\ref{radiative3}) can be a 
major supersymmetric 
background to neutralino production (~\ref{radiative1}) if  sneutrinos
decay mainly invisibly, e.g.  via $\tilde\nu\to\tilde \chi^0_1\nu$.
This leads to so called ``virtual LSP'' scenario~\cite{Datta:1996ur}.  
However, if kinematically allowed, other visible decay channels 
like $\tilde\nu\to\tilde\chi^\pm_1\ell^\mp$ reduce the background rate 
from radiative sneutrino production. For example in the SPS~1a 
scenario~\cite{Ghodbane:2002kg, Allanach:2002nj} of the MSSM we have 
${\rm  BR}(\tilde\nu_e\to\tilde\chi_1^0\nu_e)=85\%$.

Furthermore,  neutralino production $e^+e^- \to \tilde\chi_1^0
\tilde\chi^0_2$ followed by subsequent radiative neutralino
decay~\cite{Haber:1988px} $\tilde\chi^0_2 \to \tilde\chi^0_1 \gamma$
is also a potential background.  However, significant branching ratios
${\rm BR}(\tilde\chi^0_2 \to \tilde\chi^0_1 \gamma)>10\%$ are 
obtained only for small values of $\tan\beta<5$ and/or $M_1\sim
M_2$~\cite{Ambrosanio:1995it,Ambrosanio:1995az,Ambrosanio:1996gz}.
Thus,  we neglect this background, detailed  discussions
of which  can be found in Refs.~\cite{Ambrosanio:1995az,Ambrosanio:1996gz,
Baer:2002kv}.
\section{The Effect of Longitudinal Beam Polarisation}
\label{sec:pol}
In this Section we  study in detail the effect of 
longitudinal beam polarization on the cross section for radiative
neutralino production in NMSSM, and MSSM, in electron-positron
collisions. We shall also study
the beam polarization dependence of the  background processes.
The  cross section with polarized $e^\pm$ beams, with polarization 
$P_{e^\pm}$~($|P_{e^\pm}| \le 1$), can be written as 
\begin{eqnarray}
\sigma(P_{e^-},P_{e^+}) & = & 2\sum_{\lambda}
  \Big(\frac{1+P_{e^-}\lambda}{2}\Big)\Big(\frac{1-P_{e^+}\lambda}{2}\Big)\,
   \sigma_{\lambda},
   \label{polarization}
   \end{eqnarray}
where $\sigma_{\lambda}$ is the  helicity dependent cross
section, and where $\lambda$ is the helicity~($\lambda = \pm 1$)
of the initial particles~($e^\pm$). The signal radiative neutralino
production cross section can be enhanced, and the background can be reduced
by using a positively-polarized $e^-$ beam~($P_{e^-} > 0$)
and a negatively-polarized $e^+$ beam~($P_{e^+} < 0$).

In the NMSSM, and MSSM, the  radiative neutralino production process
proceeds proceeds mainly via the exchange of right selectrons
$\tilde e_R$. This is because, for the parameter choices that we use,
the neutralino has a significant bino component in these models, 
and the coupling to the
right selectron is significantly larger as compared to the
left selectrons $\tilde e_L$. This leads to the contribution from the right
selectron exchange to the cross section which is an order of magnitude 
larger as compared to the left selectron exchange. Furthermore, 
compared to the right selectron exchange, the
the contribution from left selectron exchange is  suppressed
due to the fact that $m_{\tilde e_R} < m_{\tilde e_L}$.
We also note that the $Z$ boson exchange contribution
to the neutralino production process is negligible
in these models. The SM background radiative neutrino process
proceeds mainly via the exchange of $W$ bosons. This means that 
positive electron beam polarization $P_{e^-}$ and negative
positron beam polarization $P_{e^+}$ will enhance the 
signal cross section, and at the same time 
reduce the background~\cite{Choi:1999bs}.
We note that $|P_{e^-}| > 0.8$ and $|P_{e^+}| > 0.5$
are designed at the International Linear 
Collider~\cite{:2007sg}.

A quantitative measure of the  excess of photons from the 
radiative neutralino production over the SM background
photons is  the theoretical significance defined as 
\begin{eqnarray}
S & = & \frac {N_S}{\sqrt{(N_S + N_B)}} 
   =  \frac{\sigma_{\mathrm{S}}}{\sqrt{\sigma_{\mathrm{S}}
     +\sigma_{\mathrm{B}}}} \sqrt{\mathcal{L}},
\label{significance}
\end{eqnarray}
where $N_S$ and $N_B$ define the number of signal and background events, 
respectively,  $\sigma_{\mathrm{S}}$ and $\sigma_{\mathrm{B}}$ 
are the respective cross sections, and 
${\mathcal{L}}$ is the integrated luminosity.  
If the  theoretical significance has a value  $S = 1$ for a 
signal, then that  signal can be measured
at a  $68\%$ confidence level. In addition, we must also consider
the signal to background ratio
\begin{eqnarray}
r &= & \frac{\sigma_{\mathrm{S}}}{\sigma_{\mathrm{B}}}\enspace .
\label{sigbck}
\end{eqnarray}
For a signal to be detectable at the International Linear Collider
we must have $ S > 1 \quad \mathrm{and}\quad r > 1\%\,.$
These estimates are expected to be  rough estimates which 
will enable us to 
decide whether an excess of signal photons can be measured 
over the background photons. A detailed  Monte Carlo  analysis
is beyond the scope of the present paper.


\begin{table}[t!]
\setlength{\belowcaptionskip}{11pt}
\centering
\caption{  Cross sections $\sigma$, significance $S$, and signal to 
background ratio $r$ for different beam polarizations $(P_{e^-}|
P_{e^+})$ for MSSM SPS~1a  at $\sqrt{s} = 350\GeV$, with
$\tan\beta=10$ and $\mathcal{L}=500\fb^{-1}$.}
\label{tab:S350}
\begin{tabular}{c|ccccccc}
\hline
 $(P_{e^-}|P_{e^+})$ &$(0|0)$&$(0.8|0)$&$(0.8|-0.3)$&$(0.8|-0.6)$&$(0.9|0)$&$(0.9|-0.3)$&$(0.9|-0.6)$\\[1mm]
\hline
$\sigma(\signal)$&$20\fb$&$35\fb$&$46\fb$&$56.4\fb$&$37\fb$&$48.4\fb$&$60\fb$\\[1mm]
$\sigma(\Backgroundnu)$& $2991\fb$ & $793\fb$ & $685\fb$ & $579\fb$&$518\fb$&$501\fb$&$484\fb$\\[1mm] 
$S$ & $8$ &$27$ & $38$ & $50$&$35$& $46$&$57.5$\\[1mm]
$r$ & $0.7\%$ & $4.4\%$ & $6.7\%$ & $9.7\%$&$7.1\%$&$9.6\%$&$12.4\%$\\[1mm]
\hline 
\end{tabular}
\end{table} 
\medskip
\begin{table}[t!]
\setlength{\belowcaptionskip}{11pt}
\centering
\caption{  Cross sections $\sigma$, significance $S$, and signal to 
background ratio $r$ for different beam polarizations $(P_{e^-}|
P_{e^+})$ for MSSM SPS~1a  at $\sqrt{s} = 500\GeV$, with
$\tan\beta=10$ and $\mathcal{L}=500\fb^{-1}$.}
\label{tab:S500}
\begin{tabular}{c|ccccccc}
\hline
 $(P_{e^-}|P_{e^+})$ &$(0|0)$&$(0.8|0)$&$(0.8|-0.3)$&$(0.8|-0.6)$&$(0.9|0)$&$(0.9|-0.3)$&$(0.9|-0.6)$\\[1mm]
\hline
$\sigma(\signal)$&$23\fb$&$40\fb$&$52\fb$&$64\fb$&$42\fb$&$55\fb$&$67\fb$\\[1mm]
$\sigma(\Backgroundnu)$& $2942\fb$ & $597\fb$ & $423\fb$ & $250\fb$&$304\fb$&$218\fb$&$133\fb$\\[1mm] 
$S$ & $9.4$ &$35.4$ & $53.3$ & $80.8$&$50.5$& $74.4$&$106$\\[1mm]
$r$ & $0.8\%$ & $6.7\%$ & $12.3\%$ & $25.6\%$&$13.8\%$&$25.2\%$&$50.4\%$\\[1mm]
\hline 
\end{tabular}
\end{table} 
\begin{table}[t!]
\setlength{\belowcaptionskip}{11pt}
\centering
\caption{  Cross sections $\sigma$, significance $S$, and signal to 
background ratio $r$ for different beam polarizations $(P_{e^-}|
P_{e^+})$ for MSSM SPS~1a  at $\sqrt{s} = 650\GeV$, with
$\tan\beta=10$ and $\mathcal{L}=500\fb^{-1}$.}
\label{tab:S650}
\begin{tabular}{c|ccccccc}
\hline
 $(P_{e^-}|P_{e^+})$ &$(0|0)$&$(0.8|0)$&$(0.8|-0.3)$&$(0.8|-0.6)$&$(0.9|0)$&$(0.9|-0.3)$&$(0.9|-0.6)$\\[1mm]
\hline
$\sigma(\signal)$&$18\fb$&$32\fb$&$42\fb$&$52\fb$&$34\fb$&$44\fb$&$55\fb$\\[1mm]
$\sigma(\Backgroundnu)$& $2869\fb$ & $578\fb$ & $408\fb$ & $237\fb$&$292\fb$&$207\fb$&$123\fb$\\[1mm] 
$S$ & $7.5$ &$29$ & $44$ & $68$&$42$& $62$&$92$\\[1mm]
$r$ & $0.63\%$ & $5.5\%$ & $10.3\%$ & $22\%$&$11.6\%$&$21.3\%$&$44.7\%$\\[1mm]
\hline 
\end{tabular}
\end{table} 

We have studied  the radiative production of lightest neutralino 
for three different electron-positron 
center of mass energies, namely  $\sqrt{s} =350$ GeV,
$500$ GeV, and $ 650$ GeV, respectively with longitudinally polarized beams
for the three models---MSSM SPS1a, MSSM EWSB and NMSSM using CALCHEP.
For MSSM SPS 1a, we have calculated the 
beam polarization dependence of the signal 
cross section $\sigma(\signal)$ and the background 
cross section $\sigma(\Backgroundnu)$,
the significance S and signal to background ratio $r$ at  $\sqrt{s} = 350$ GeV,
$ 500$ GeV  and $ 650$ GeV  for the input parameters
as in  Table ~\ref{parMSSM}. We report the values of cross-sections 
$\sigma(\signal)$ and $\sigma(\Backgroundnu)$, $S$ and $r$ for a
specific set of beam of polarizations 
$(P_{e^- }|P_{e^+}) = (0|0)$, $(0.8|0)$, $(0.8|-0.3)$, $(0.8|-0.6)$, 
$(0.9|0)$, $(0.9|-0.3)$, and $(0.9|-0.6)$ in Tables~\ref{tab:S350}, 
~\ref{tab:S500}, and ~\ref{tab:S650}, respectively. 
We observe that there is a 
large enhancement in  the value of $ S = 106 $ and $ r = 50.4 \%$ 
for $(P_{e^- }|P_{e^+}) =  (0.9|-0.6)$ compared to all other 
chosen set of polarization values for  $\sqrt{s} = 500$ GeV.
For a previous study of the beam polarization effects for the
radiative neutralino production in  MSSM,
see Ref. \cite{Dreiner:2007vm}.
\medskip

Similarly, in the Tables~\ref{tab:M350}, ~\ref{tab:M500}, and ~\ref{tab:M650},
we discuss the values of cross-sections $\sigma(\signal)$ 
and $\sigma(\Backgroundnu)$, $S$ and $ r$ for a 
specific set of beam  polarization values 
$(P_{e^- }|P_{e^+}) = (0|0)$, $(0.8|0)$, $(0.8|-0.3)$, 
$(0.8|-0.6)$, $(0.9|0)$,
$(0.9|-0.3)$, and $(0.9|-0.6)$ for the case of MSSM EWSB.
There is again a significant  increase  for the value of 
$S=12.5$ and $r = 5 \%$ for $(P_{e^- }|P_{e^+}) =  (0.9|-0.6)$ 
compared to other chosen polarization values for  $\sqrt{s} =500$~GeV.
However, the enhancement is relatively samll as  compared to the
MSSM SPS 1a scenario.

\begin{table}[b!]
\setlength{\belowcaptionskip}{11pt}
\centering
\caption{  Cross sections $\sigma$, significance $S$, and signal to 
background ratio $r$ for different beam polarizations $(P_{e^-}|
P_{e^+})$ for MSSM EWSB  at $\sqrt{s} = 350\GeV$, with
$\tan\beta=10$ and $\mathcal{L}=500\fb^{-1}$.}
\label{tab:M350}
\begin{tabular}{c|ccccccc}
\hline
 $(P_{e^-}|P_{e^+})$ &$(0|0)$&$(0.8|0)$&$(0.8|-0.3)$&$(0.8|-0.6)$&$(0.9|0)$&$(0.9|-0.3)$&$(0.9|-0.6)$\\[1mm]
\hline
$\sigma(\signal)$&$1.7\fb$&$2.9\fb$&$3.8\fb$&$4.7\fb$&$3.1\fb$&$4\fb$&$4.9\fb$\\[1mm]
$\sigma(\Backgroundnu)$& $2991\fb$ & $793\fb$ & $685\fb$ & $579\fb$&$518\fb$&$501\fb$&$484\fb$\\[1mm]  
$S$ & $0.7$ &$2.3$ & $3.2$ & $4.3$&$3$& $4$&$5$\\[1mm]
$r$ & $0.06\%$ & $0.4\%$ & $0.56\%$ & $0.81\%$&$0.6\%$&$0.8\%$&$1.02\%$\\[1mm]
\hline 
\end{tabular}
\end{table} 
\begin{table}[h!]
\setlength{\belowcaptionskip}{11pt}
\centering
\caption{  Cross sections $\sigma$, significance $S$, and signal to 
background ratio $r$ for different beam polarizations $(P_{e^-}|
P_{e^+})$ for MSSM EWSB   at $\sqrt{s} = 500\GeV$, with
$\tan\beta=10$ and $\mathcal{L}=500\fb^{-1}$.}
\label{tab:M500}
\begin{tabular}{c|ccccccc}
\hline
 $(P_{e^-}|P_{e^+})$ &$(0|0)$&$(0.8|0)$&$(0.8|-0.3)$&$(0.8|-0.6)$&$(0.9|0)$&$(0.9|-0.3)$&$(0.9|-0.6)$\\[1mm]
\hline
$\sigma(\signal)$&$2.2\fb$&$3.9\fb$&$5.1\fb$&$6.3\fb$&$4.1\fb$&$5.4\fb$&$6.6\fb$\\[1mm]
$\sigma(\Backgroundnu)$& $2942\fb$ & $597\fb$ & $423\fb$ & $250\fb$&$804\fb$&$218\fb$&$133\fb$\\[1mm] 
$S$ & $0.9$ &$3.6$ & $5.5$ & $8.8$&$5.2$& $8$&$12.5$\\[1mm]
$r$ & $0.075\%$ & $0.65\%$ & $1.2\%$ & $2.5\%$&$1.3\%$&$2.5\%$&$5\%$\\[1mm]
\hline 
\end{tabular}
\end{table} 
\begin{table}[h!]
\setlength{\belowcaptionskip}{11pt}
\centering
\caption{  Cross sections $\sigma$, significance $S$, and signal to 
background ratio $r$ for different beam polarizations $(P_{e^-}|
P_{e^+})$ for MSSM EWSB  at $\sqrt{s} = 650\GeV$, with
$\tan\beta=10$ and $\mathcal{L}=500\fb^{-1}$.}
\label{tab:M650}
\begin{tabular}{c|ccccccc}
\hline
 $(P_{e^-}|P_{e^+})$ &$(0|0)$&$(0.8|0)$&$(0.8|-0.3)$&$(0.8|-0.6)$&$(0.9|0)$&$(0.9|-0.3)$&$(0.9|-0.6)$\\[1mm]
\hline
$\sigma(\signal)$&$1.9\fb$&$3.4\fb$&$4.4\fb$&$5.4\fb$&$3.5\fb$&$4.6\fb$&$5.7\fb$\\[1mm]
$\sigma(\Backgroundnu)$& $2869\fb$ & $578\fb$ & $408\fb$ & $237\fb$&$292\fb$&$207\fb$&$123\fb$\\[1mm] 
$S$ & $0.8$ &$3$ & $4.8$ & $7.7$&$4.5$& $7$&$11$\\[1mm]
$r$ & $0.06\%$ & $0.6\%$ & $1\%$ & $2.3\%$&$1.2\%$&$2.2\%$&$4.6\%$\\[1mm]
\hline 
\end{tabular}
\end{table} 

Finally, in 
Tables ~\ref{tab:N350}, \ref{tab:N500}, \ref{tab:N650}, we show the results 
for the case of NMSSM. We get the similar pattern but the value of 
$S=7$ and $r = 2.8 \%$ for $(P_{e^- }|P_{e^+}) =  (0.9|-0.6)$  is considerably
smaller as  compared to  that of 
$ S = 106$ and $ r = 50.4 \%$  in  MSSM SPS 1a at  $\sqrt{s} = 500 \GeV$. 
Thus, if the radiative production of neutralinos is observed at a
linear collider with polarized beams, then on the basis of the  
observed event rate it may be
possible to distinguish between MSSM and NMSSM as the underlying low energy
supersymmetric model. 

\begin{table}[h!]
\setlength{\belowcaptionskip}{11pt}
\centering
\caption{  Cross sections $\sigma$, significance $S$, and signal to 
background ratio $r$ for different beam polarizations $(P_{e^-}|
P_{e^+})$ for NMSSM  at $\sqrt{s} = 350\GeV$, with
$\tan\beta=10$ and $\mathcal{L}=500\fb^{-1}$.}
\label{tab:N350}
\begin{tabular}{c|ccccccc}
\hline
 $(P_{e^-}|P_{e^+})$ &$(0|0)$&$(0.8|0)$&$(0.8|-0.3)$&$(0.8|-0.6)$&$(0.9|0)$&$(0.9|-0.3)$&$(0.9|-0.6)$\\[1mm]
\hline
$\sigma(\signal)$&$1.1\fb$&$2\fb$&$2.6\fb$&$3.2\fb$&$2.1\fb$&$2.7\fb$&$3.3\fb$\\[1mm]
$\sigma(\Backgroundnu)$& $2991\fb$ & $793\fb$ & $685\fb$ & $579\fb$&$518\fb$&$501\fb$&$484\fb$\\[1mm] 
$S$ & $0.45$ &$1.6$ & $2.2$ & $3$&$2$& $2.7$&$3.3$\\[1mm]
$r$ & $0.04\%$ & $0.25\%$ & $0.4\%$ & $0.55\%$&$0.4\%$&$0.54\%$&$0.68\%$\\[1mm]
\hline 
\end{tabular}
\end{table} 
\begin{table}[h!]
\setlength{\belowcaptionskip}{11pt}
\centering
\caption{  Cross sections $\sigma$, significance $S$, and signal to 
background ratio $r$ for different beam polarizations $(P_{e^-}|
P_{e^+})$ for NMSSM  at $\sqrt{s} = 500\GeV$, with
$\tan\beta=10$ and $\mathcal{L}=500\fb^{-1}$.}
\label{tab:N500}
\begin{tabular}{c|ccccccc}
\hline
 $(P_{e^-}|P_{e^+})$ &$(0|0)$&$(0.8|0)$&$(0.8|-0.3)$&$(0.8|-0.6)$&$(0.9|0)$&$(0.9|-0.3)$&$(0.9|-0.6)$\\[1mm]
\hline
$\sigma(\signal)$&$1.2\fb$&$2.2\fb$&$2.8\fb$&$3.5\fb$&$2.3\fb$&$3\fb$&$3.7\fb$\\[1mm]
$\sigma(\Backgroundnu)$& $2942\fb$ & $597\fb$ & $423\fb$ & $250\fb$&$804\fb$&$218\fb$&$133\fb$\\[1mm] 
$S$ & $0.5$ &$2$ & $3$ & $5$&$3$& $4.5$&$7$\\[1mm]
$r$ & $0.04\%$ & $0.4\%$ & $0.66\%$ & $1.4\%$&$0.76\%$&$1.4\%$&$2.8\%$\\[1mm]
\hline 
\end{tabular}
\end{table} 
\begin{table}[h!]
\setlength{\belowcaptionskip}{11pt}
\centering
\caption{  Cross sections $\sigma$, significance $S$, and signal to 
background ratio $r$ for different beam polarizations $(P_{e^-}|
P_{e^+})$ for NMSSM  at $\sqrt{s} = 650\GeV$, with
$\tan\beta=10$ and $\mathcal{L}=500\fb^{-1}$.}
\label{tab:N650}
\begin{tabular}{c|ccccccc}
\hline
 $(P_{e^-}|P_{e^+})$ &$(0|0)$&$(0.8|0)$&$(0.8|-0.3)$&$(0.8|-0.6)$&$(0.9|0)$&$(0.9|-0.3)$&$(0.9|-0.6)$\\[1mm]
\hline
$\sigma(\signal)$&$1\fb$&$1.8\fb$&$2.3\fb$&$2.8\fb$&$1.9\fb$&$2.4\fb$&$3\fb$\\[1mm]
$\sigma(\Backgroundnu)$& $2869\fb$ & $578\fb$ & $408\fb$ & $237\fb$&$292\fb$&$207\fb$&$123\fb$\\[1mm] 
$S$ & $0.42$ &$1.7$ & $2.5$ & $4$&$2.5$& $3.7$&$6$\\[1mm]
$r$ & $0.03\%$ & $0.3\%$ & $0.56\%$ & $1.2\%$&$0.65\%$&$1.16\%$&$2.44\%$\\[1mm]
\hline 
\end{tabular}
\end{table} 

\section{Summary and Conclusions}
\label{sec:conclusions}
The nonminimal supersymmetric standard model~(NMSSM)
solves $\mu$ problem of MSSM  in an elegant manner and is, thus, an attractive
alternative to the MSSM. We have carried out a detailed study of 
the radiative neutralino production process
$e^+e^- \to\tilde\chi^0_1 \tilde\chi^0_1\gamma$ in NMSSM for 
ILC energies,  and
compared the results  with the corresponding results in 
the MSSM for both unpolarized and polarized  $e^-$ and $e^+$ beams.
This process has a signature of a high energy photon and missing energy.
We have  obtained a typical set of parameter values for the NMSSM by imposing 
theoretical and experimental constraints on the parameter space of NMSSM.
For the  set of parameter values that we obtain in this manner, 
the lightest neutralino in NMSSM
has a significant admixture of the fermionic component of the singlet chiral
superfield $S$. Using this parameter set, we have studied in detail  the 
radiative neutralino production cross section in NMSSM for the 
ILC energies for  both unpolarized and polarized $e^-$ and $e^+$ beams. 
For comparison with MSSM, we have used the  MSSM SPS~1a and MSSM~EWSB models.
The background to this process comes from the SM  process  
$e^+e^- \to \nu \bar\nu \gamma$,
as well as the background from the supersymmetric process
$e^+e^- \to \tilde\nu \tilde\nu^\ast \gamma$. All these processes have 
a signature of a highly energetic photon with missing energy.
The photon energy distribution $d\sigma/dE_{\gamma}$, 
and the total cross section as  a function of the total energy 
have been calculated for the NMSSM and  for  MSSM SPS~1a scenario 
at $\sqrt{s} = 500$ GeV using  CALCHEP package.  
Because of the admixture of a singlet in the lightest
neutralino, the cross section as a function of energy 
for the radiative neutralino production is lower in NMSSM than in  MSSM.
We have also studied the dependence of the cross section for radiative
neutralino production on the $SU(2)_L$ gaugino mass parameter 
$M_2$ and the Higgs(ino) mass parameter $\mu$, 
as well as its dependence on the selectron~($\tilde e_R, \tilde e_L$)
masses in NMSSM, and compared it with the corresponding results in
MSSM. In order to quantify whether an excess of signal 
photons, $N_{\mathrm{S}}$, can be measured over the background
photons, $N_{\rm B}$, from radiative neutrino production, we have
analyzed the theoretical statistical significance 
$S = N_{\rm S}/\sqrt{N_{\rm S} + N_{\rm B}}$.
At the ILC, electron and positron beam polarizations can be used
to significantly enhance the signal and suppress the background
simultaneously.  We have shown that the significance can then be
increased almost by an order of magnitude, 
e.g. with $(P_{e^-},P_{e^+}) = (0.8,-0.6)$ 
compared to $(P_{e^-},P_{e^+})=(0,0)$. 
A possible feed-back of ILC$_{500}$ results 
could motivate the immediate use of the low-luminosity option of
the ILC  at $\sqrt{s}=650$~GeV in order to resolve model ambiguities
between NMSSM and MSSM  even at an early stage of
the experiment and outline future search strategies at the
upgraded ILC at 1 TeV.
In our scenarios, the
signal cross section for $(P_{e^-}|P_{e^+}) = (0.8|-0.6)$ is larger
than $3.5\fb$, the significance $S>5$, and the signal to background
ratio is about $r > 1\%$. The background cross section can be reduced to
$250\fb$.  Increasing the positron beam polarization to $P_{e^+} =
-0.6$, both the signal cross section and the significance 
increase significantly. 
Thus the electron and positron beam
polarization at the ILC are  essential tools to observe radiative
neutralino production.
\section{Acknowledgements}
P.~N.~P. would like to thank 
The Institute of Mathematical Sciences, Chennai for hospitality
where was this work was initiated. The work of P.~N.~P. is supported by the
Council of Scientific and Industrial Research, India, and by the
J. C. Bose National Fellowship of the Department of Science and Technology, 
India. P.~N.~P. would like to thank the Inter University Centre
for Astronomy and Astrophysics for hospitality while this work was completed.
\newpage
\begin{appendix}
\section{Superpotential, neutralino mass matrix and couplings}
\label{appendix: superpotential and couplings}
For completeness we  summarize here
the couplings of the neutralinos to electrons and 
the scalar partners of electrons, the selectrons,  in MSSM and in NMSSM.  
These couplings can be obtained from the 
neutralino mixing matrix. To obtain the neutralino mixing matrix for the MSSM, 
we recall that the neutralino mass matrix obtains contributions from 
part of the MSSM superpotential 
\begin{eqnarray}
W_{\mathrm{MSSM}} & = & \mu H_1 H_2,
\label{WMSSM}
\end{eqnarray}
where $H_1$ and $H_2$ are the two Higgs doublet chiral superfields,
and $\mu$ is the supersymmetric Higgs(ino) parameter. In addition
to the contribution from the superpotential, the neutralino mass  matrix
receives contributions from the interactions between gauge and  matter
multiplets, as well as contributions from the soft supersymmetry breaking 
masses for the gauginos. Including all these contributions, the neutralino mass 
matrix, in the bino, wino, higgsino basis 
$(-i\lambda', -i\lambda^3, \psi_{H_1}^1,
\psi_{H_2}^2)$ can be written as~\cite{Bartl:1989ms, Haber:1984rc}
\begin{eqnarray}
\label{mssmneut}
M_{\mathrm{MSSM}} =
\begin{pmatrix}
M_1 & 0   & - m_Z \sw \cos\beta & \phantom{-}m_Z\sw \sin\beta \\
0   & M_2 & \phantom{-} m_Z \cw \cos\beta  & -m_Z \cw\sin\beta \\
 - m_Z \sw \cos\beta &\phantom{-} m_Z \cw \cos\beta  & 0 & -\mu\\
\phantom{-}m_Z\sw \sin\beta& -m_Z \cw\sin\beta & -\mu & 0
\end{pmatrix},
\end{eqnarray}
where $M_1$ and $M_2$ are the $U(1)_Y$ and the $SU(2)_L$
soft supersymmetry breaking gaugino mass parameters, respectively, and
$\tan\beta = v_2 /v_1$ is the ratio of the vacuum expectation
values of the neutral components of the two Higgs doublet 
fields $H_1$ and $H_2$, respectively. Furthermore,
$m_Z$ is the $Z$ boson mass, and $\theta_w$ is the
weak mixing angle. We shall assume that all the parameters
in the matrix  $M$  are real, in which case 
$M$ is a real 
symmetric matrix and can be diagonalised by an orthogonal matrix. 
If one of the  eigenvalues of  $M$ is negative, 
one can diagonalize  this matrix using a  unitary matrix $N$, the neutralino 
mixing matrix, to get a positive  diagonal 
matrix~\cite{Haber:1984rc}:  
\begin{eqnarray}
\label{mssmdiag}
N^\ast M_{\mathrm{MSSM}} N^{-1} =   \mathrm{diag}\begin{pmatrix}m_{\chi_1^0}, 
& m_{\chi_2^0}, & m_{\chi_3^0}, & m_{\chi_4^0} \end{pmatrix}.
\end{eqnarray}
where  $m_{\chi_i^0}~(i = 1, 2, 3, 4)$ are neutralino masses arranged in order 
of increasing value.

For the NMSSM, the relevant part of the superpotential is
\begin{eqnarray}
\label{WNMSSM}
W_{\mathrm{NMSSM}} & = &  \lambda S H_1 H_2 - \frac{\kappa}{3}S^3,
\end{eqnarray}
where $S$ is the Higgs singlet chiral superfield.
In the basis   $(-i\lambda', -i\lambda^3, \psi_{H_1}^1,
\psi_{H_2}^2, \psi_S)$, the neutralino mass matrix for the NMSSM 
can then be written as~\cite{pnp1, pnp2}
\begin{eqnarray}
\label{nmssmneut}
M_{\mathrm{NMSSM}} = \begin{pmatrix}
M_1 & 0   & - m_Z \sw \cos\beta & \phantom{-}m_Z\sw \sin\beta & 0 \\
0   & M_2 & \phantom{-} m_Z \cw \cos\beta  & -m_Z \cw\sin\beta & 0 \\
- m_Z \sw \cos\beta &\phantom{-} m_Z \cw \cos\beta  & 0 & -\lambda x 
& -\lambda v_2\\
\phantom{-}m_Z\sw \sin\beta& -m_Z \cw\sin\beta & -\lambda x & 0 & -\lambda v_1\\
0 & 0 &  -\lambda v_2 & -\lambda v_1 & 2 \kappa x
\end{pmatrix},
\end{eqnarray}
where $<S> = x$ is the vacuum expectation value of the singlet Higgs field.
As in the case of MSSM, we can use a unitary matrix $N'$ to get
a positive semidefinite diagonal matrix with the neutralino masses
$m_{\chi_i^0}~(i = 1, 2, 3, 4, 5)$~\cite{pnp1, pnp2}:
\begin{eqnarray}
\label{nmssmdiag}
N'^\ast M_{\mathrm{NMSSM}} N'^{-1} = 
\mathrm{diag}\begin{pmatrix}m_{\chi_1^0}, & m_{\chi_2^0}, & m_{\chi_3^0},
& m_{\chi_4^0} &  m_{\chi_5^0} \end{pmatrix}.
\end{eqnarray}
The Lagrangian for the interaction of neutralinos, electrons, selectrons and 
$Z$ bosons for MSSM is given by~\cite{Haber:1984rc}
\begin{eqnarray}
{\mathcal L} &=& (- \frac {\sqrt{2}e}{\cw} N_{11}^*)
                    \bar{f}_eP_L\tilde{\chi}^0_1\tilde{e}_R
                 + \frac{e}{\sqrt{2} \sw} (N_{12} + \tw N_{11}) 
                   \bar{f}_e P_R\tilde{\chi}^0_1\tilde{e}_L \nonumber \\
     & & + \frac{e}{4 \sw \cw} \left(|N_{13}|^2 - |N_{14}|^2\right)
           Z_\mu \bar{\tilde{\chi}}_1^0\gamma^\mu \gamma^5\tilde{\chi}_1^0
           \nonumber \\
          && + e Z_\mu \bar{f}_e \gamma^\mu 
             \big[ \frac{1}{\sw\cw}\left(\frac{1}{2} - \sw[2]\right) P_L 
                     - \tw  P_R\big] {f}_e + \mathrm{h. c.},
\label{mssmlagrangian}
\end{eqnarray} 
with the electron, selectron, neutralino and $Z$ boson fields denoted by
$f_e$, $\tilde{e}_{L,R}$, $\tilde{\chi}_1^0$, and $Z_\mu$, respectively,  
and $P_{R, L} = \frac{1}{2} \left(1 \pm \gamma^5\right)$. The corresponding
interaction Lagrangian for NMSSM is obtained from (~\ref{mssmlagrangian})
by replacing $N_{ij}$ with  $N'_{ij}.$  The different  vertices 
following from (~\ref{mssmlagrangian}) are shown in Table~\ref{feynmandiag}.
The  couplings of the lightest neutralino to electrons, selectrons and $Z$ 
boson are determined by the corresponding elements of the neutralino mixing
matrix~($N_{ij}$  or $N'_{ij}$). 
\begin{table}[h!]
\begin{center}
\caption{Vertices corresponding to various terms in the interaction 
Lagrangian (~\ref{mssmlagrangian}) for MSSM. In addition we have 
also shown the vertices for selectron-photon  and electron-photon
interactions. The vertices for the NMSSM 
are obtained by replacing $N_{ij}$ with $N'_{ij}$.}
\vspace{5mm}
\begin{tabular}{lccccccccl}
\hline
\\
Vertex & & & & & & & & & Vertex Factor\\
& & &  & &&& &&\\
\hline
& & & & & && &&\\
right~selectron - electron - neutralino 
& & & & & && && {$\frac {-i e \sqrt{2}}{\cw}N_{11}^* P_L$}\\
& & & & & && &&\\
left~selectron - electron - neutralino
& & & & & && &&{$\frac{i e}{\sqrt{2} \sw} (N_{12} + \tw N_{11}) P_R$}\\
& & & & & && &&\\
neutralino - $Z^0$ - neutralino
& & & & & && && {$ \frac{i e}{4 \sw \cw} 
           \left(|N_{13}|^2 - |N_{14}|^2\right) \gamma^\mu \gamma^5$}\\
& & & & & && && \\
electron - $Z^0$ - electron
& & & & & && && {$ i e \gamma^\mu \big[ \frac{1}{\sw\cw}\left(\frac{1}{2} 
               - \sw[2]\right) P_L - \tw  P_R\big] $} \\
& & & & & && &&\\
selectron - photon - selectron
& & & & & && && {$ i e (p_1 + p_2)^\mu$}\\
& & & & & && && \\
electron - photon - electron
& & & & & && && {$ i e \gamma^\mu$}\\
& & & & &  && && \\
\\
\hline
\bottomrule
\end{tabular}
\label{feynmandiag}
\end{center}
\end{table}
\end{appendix}
%


\end{document}